%% file: main.tex
\documentclass[
 twocolumn,
 aps,
 prl,
 10pt,
 superscriptaddress,
 reprint,
 bibnotes,
 amsmath,amssymb,
 floatfix
]{revtex4-2}

\usepackage{times}
\usepackage{stmaryrd}
\usepackage[normalem]{ulem}
\usepackage{comment}
\usepackage[usenames,dvipsnames]{color}
\usepackage{color}
\usepackage[pdftex]{graphicx}%arxiv;pdf
\usepackage{dcolumn}
\usepackage{bm}
\usepackage{here}
\usepackage{comment}
\usepackage{siunitx}
\usepackage[colorlinks,linkcolor=blue,citecolor=blue,urlcolor=black]{hyperref}

\include{commands}

\begin{document}
\title{Optimal Finite-time Maxwell's Demons in Langevin Systems}
\author{Takuya Kamijima}
\email{kamijima@noneq.t.u-tokyo.ac.jp}
\affiliation{
 Department of Applied Physics, The University of Tokyo, 7-3-1 Hongo, Bunkyo-ku, Tokyo 113-8656, Japan}

 \author{Asuka Takatsu}
 \affiliation{Department of Mathematical Sciences, Tokyo Metropolitan University, 1-1 Minami-osawa, Hachioji-shi, Tokyo, Japan}
 \affiliation{RIKEN Center for Advanced Intelligence Project (AIP), 1-4-1 Nihonbashi, Chuo-ku, Tokyo, Japan}

\author{Ken Funo}
\affiliation{
 Department of Applied Physics, The University of Tokyo, 7-3-1 Hongo, Bunkyo-ku, Tokyo 113-8656, Japan}

\author{Takahiro Sagawa}
\affiliation{
 Department of Applied Physics, The University of Tokyo, 7-3-1 Hongo, Bunkyo-ku, Tokyo 113-8656, Japan}
\affiliation{
 Quantum-Phase Electronics Center (QPEC), The University of Tokyo, 7-3-1 Hongo, Bunkyo-ku, Tokyo 113-8656, Japan}

\begin{abstract}
We identify the optimal protocols to achieve the minimal entropy production in finite-time information exchange processes in Langevin systems, on the basis of optimal transport theory.
Our general results hold even for non-Gaussian cases, while we derive a concise expression of the minimal entropy production for Gaussian processes.
In particular, we apply our results to Maxwell's demons that perform measurement and feedback, and demonstrate Gaussian and non-Gaussian models of optimal demons operating in finite time.
Our results provide a general strategy for controlling Langevin systems, including colloidal particles and biomolecules, in a thermodynamically optimal manner beyond the quasi-static limit.
\end{abstract}
%comment: 今の設定だと始・終分布だけ固定なので、始・終ポテンシャルも固定しないとminimum workは定まらない
%単数複数チェック 
%We reveal that the optimal feedback protocol requires the nonconservative force in general.
%, showing that the correlation coefficient quantifies the tradeoff relation between partial entropy productions.

\maketitle
\textit{Introduction.---}
The concept of Maxwell's demon dates back to the nineteenth century, with its modern formulation being developed within the framework of stochastic thermodynamics, where the demon is formulated to perform measurement and feedback on thermodynamic systems \cite{leff2002maxwell,Parrondo-Horowitz-Sagawa2015}.
The fundamental thermodynamic costs of the demon's operations, such as measurement, feedback and information erasure, have been revealed as an extension of the second law of thermodynamics incorporating information \cite{Sagawa-Ueda2008quantumfeedback,Sagawa-Ueda2009erasure,Esposito2011Landauer,Sagawa-Ueda2012-FT,sagawaNJP2013,ItoSagawa2013causal,infoflow-Holowitz-2014,hartich2014transfer} and demonstrated experimentally \cite{toyabe-2010exp-Jarzynski,berut2012Landauer_exp,Koski-Sagawa2014experiment,Vidrighin-2016photonic-demon,Nathanael-2017quatum-demon,masuyama-2018demon-SC,Ribezzi-Marco2019exp-Maxwell}.
%where $W$ is the extracted work, $\calF_X$ is the nonequilibrium free energy of the engine $X$ \cite{Esposito2011Landauer,Parrondo-Horowitz-Sagawa2015}, $T$ is the temperature of the environment (the Boltzmann constant is set to unity), and $\Delta I$ represents the mutual information exchange induced by the demon.
%The extracted work can exceed the free energy difference when mutual information is consumed through feedback (i.e., $\Delta I<0$), as demonstrated in various experiments \cite{toyabe-2010exp-Jarzynski,Koski-Sagawa2014experiment,Vidrighin-2016photonic-demon,Nathanael-2017quatum-demon,masuyama-2018demon-SC,Ribezzi-Marco2019exp-Maxwell}.
While the optimal protocols to achieve the fundamental second-law bounds have been established \cite{Parrondo-Horowitz-Sagawa2015}, such protocols in general incorporate infinitely slow driving and are unattainable in finite time.
Despite growing interest in finite-time thermodynamic bounds such as the thermodynamic speed limits \cite{Shiraishi-Funo-Saito2018speed,Ito2018geometry,Ito-Dechant2020info,Falasco-Esposito2020dissipation-time,Lee-Park_highly2022,Hamazaki2022speed,dechant2022minimum,Van-Saito-unification2023,Sabbagh-2024-speedlimit,Ma-2020exp-scaling} and the thermodynamic uncertainty relations \cite{Barato-Seifert2015,Gingrich-Horowitz2016,Pietzonka-Barato-Seifert2016-universal-bound,Shiraishi-Saito-Tasaki2016heatengine,proesmans2017discrete,Pietzonka-Seifert2018,Hasegawa-Van2019,koyuk2020thermodynamic,Liu-Gong-Ueda2020}, the finite-time thermodynamics of information \cite{abreu-Seifert2011feedback,Proesmans2020finiteLandauer,Zhen2021boundbitreset,Nakazato-Ito2021,Taghvaei-2022-demon,Nagase-Sagawa2023infogain,FujimotoIto2023,kamijima2024discrete,Dago-2021exp-Landauer-underdamped} is yet to be established.

%とくに、\eref{quasistatic demon}の等号を達成する最適なプロトコルは知られているが、それにはinfinitely slow drivingが必須（注：クエンチが必要なのでquasi-staticとは言わない方がいいかも）。一方で、finite-timeが重要 (speed limit \cite{} や TUR \cite{} な観点からも) だが、有限時間で最適なプロトコルは未解明。
%しかし、全体のコストしかわからない。
%しかも、通常の測定やフィードバックでは片方の部分系は静止しており、その%ような状況には適応できず最小のコストは不明。

\begin{figure}
    \centering
    \includegraphics[width=0.85\linewidth]{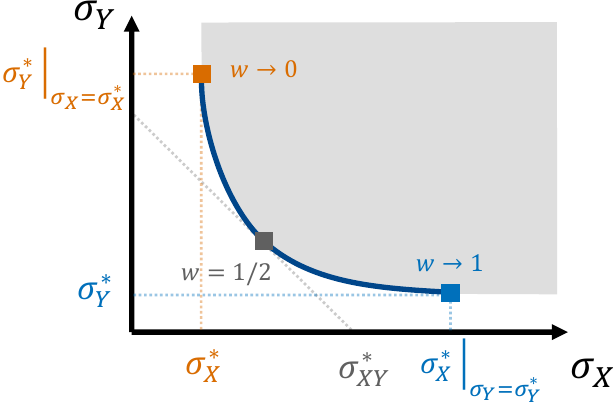}
    \caption{\label{fig: tradeoff}
    Schematic of the tradeoff relation between the partial EPs in finite time.
    The shaded region represents realizable values of the partial EPs $(\sigma_X, \sigma_Y)$, whose boundary is the blue curve representing the tradeoff relation.
    The partial EPs are minimized at the endpoints of this curve (marked by the orange and blue squares).
    %In particular, the optimal feedback process is realized at the blue square point, where $Y$ does not evolve ($\sigma_Y^* = 0$) and the EP of $X$ is optimized ($\sigma_X^* |_{\sigma_Y = \sigma_Y^*}$).
    %The inset illustrates the corresponding feedback process, where the mutual information of the demon's memory $Y$ is consumed ($\Delta I=I_o-I_f<0$) to reduce the EP of the engine $X$.
    %ガウシアンの図から以下を移動。
    The blue curve is the Pareto front of conflicting quantities $\sigma_X$ and $\sigma_Y$  (see Supplemental Material for details). In general, such Pareto front can be obtained by minimizing  $(1-w)\sigma_X + w\sigma_Y$ with sweeping the weight $0<w<1$, where taking the limit of $w\rightarrow0$ and $w\rightarrow1$ respectively yield the orange and blue squares.
   The gray square ($w=1/2$) depicts the global optimal pair $(\sigma_X, \sigma_Y)$ that achieves \eref{min sigma_XY}.
    }
\end{figure}

In this Letter, we address this issue by solving an optimization problem for the entropy production (EP) in finite-time information exchange processes in bipartite overdamped Langevin systems. 
Specifically, we consider the finite-time tradeoff relation between the EPs of two subsystems (see also \fref{fig: tradeoff} later), which provides the minimum EPs and the corresponding optimal protocols.
In view of optimal transport theory \cite{villani2003topics,villani2009}, the minimum EPs have a clear geometrical meaning and are expressed as Wasserstein distances. 
Such Wasserstein distance can be efficiently computed by utilizing monotone maps even for non-Gaussian distributions \cite{santambrogio2015OT}.
Furthermore, the minimum EPs can be expressed concisely for Gaussian cases, based solely on the correlation coefficients.

An important application of our theory is Maxwell's demon, where the subsystems play the roles of the engine and the demon's memory.
In such setups, our general results lead to the optimal energy costs for measurement and feedback in finite time, as will be demonstrated by both Gaussian and non-Gaussian models of the demons.
Moreover, we introduce a characteristic time that determines whether Maxwell's demons can be realized within a given operation time.
These results would be useful for designing various experimental systems modeled by overdamped Langevin dynamics.

%This bound is optimal in a sense that this can be achieved not only for Gaussian distributions but also for general distributions.
%Furthermore, the optimal protocol to achieve \eref{finite demon} can be obtained efficiently.
%Here, $\Lambda$ is a nonnegative quantity that depends on the initial and final distributions and can be expressed in terms of the Wasserstein distance between the conditional distributions (\eref{min sigma_X|Y general}).
%linear programmingはdiscreteだけ
%As seen from this bound, finite-time operations cause detrimental effect proportional to $\tau^{-1}$.
%In feedback processes, \eref{finite demon} indicates that one needs a finite time duration to implement Maxwell's demon.
%非保存力が要ること.

%三段楽目、figureを参照する、短め
%これらの結果は、[]のoverdampedな系に位置付けられるが、その連続極限で得ることはできず、全く異なるアプローチを必要とする。有限時間で最適なMaxwell demonを明らかにした、最適輸送を駆使.
%また、overdampedな系にはactivityの自由度による複雑な議論がなく、クリア(直感的？)な描像が存在する。

%%%%%%%%%%%%%%%%%%%%%%%%%%%%%%%%%%
\textit{Setup.---}
%In this section, we briefly review the stochastic thermodynamics for overdamped systems and optimization of the entropy production via optimal transport theory.
We consider bipartite overdamped systems, where subsystems exchange information quantified by mutual information \cite{Sagawa-Ueda2012-FT,infoflow-Holowitz-2014}.
The total system is denoted as $XY$ and the subsystems are denoted as  $X$ and $Y$. While these subsystems are regarded as the engine and the memory in the Maxwell's demon situations, we do not restrict our setup to such situations at this stage.
The total system is described by the two-dimensional spatial coordinate $\bsr=[x,y]^{\rm T}$ with $\rm T$ representing transpose, 
and the probability distribution $\pt(\bsr,t)$ follows the Fokker-Planck equation \cite{vanKampen}:
\begin{align}
    \label{Fokker-Planck}
    &\partial_t \pt = -\nabla^{\rm T}(\bsv \pt),\\
    &\bsv = -\mu(\nabla V - \bsF +  T \nabla \ln \pt),
\end{align}
where $\nabla$ represents the gradient, $\bsv = [ v_X, v_Y ]^{\rm T}$ represents the mean local velocity, and $\mu$ denotes the mobility constant.
The forces acting on the system consist of the conservative force $-\nabla V(\bsr,t)$ and the nonconservative force $\bsF(\bsr,t)$.
The total EP of the total system and the partial EPs of the subsystems are respectively given by \cite{Seifert2012,allahverdyan-2009infoflow,rosinberg02016continuous-info}
\begin{align}
    \label{sigma_XY}
    \sigma_{XY}&=\frac{1}{\mu T}\int_0^\tau dt \int d\bsr\parallel \bsv(\bsr,t)\parallel^2 \pt(\bsr,t),\\
    \label{sigma_X}
    \sigma_i&=\frac{1}{\mu T}\int_0^\tau dt\int d\bsr v_i(\bsr,t)^2 \pt(\bsr,t)  \ \ \  (i= X, Y).
\end{align}
These EPs are nonnegative in accordance with the second law and characterize energetic dissipation.
Importantly, the total EP is connected to the work and the nonequilibrium free energy of the total system, $\calF_{XY}$, through the expression $\sigma_{XY}= (-W-\Delta \calF_{XY}) / T$.
%（式を書いて(1)式と結びつける、部分エントロピー生成を書くべき？、、と思ったのですが後で出てくるから無くてもいい？微妙なところですね。全系についてだけはを書いてもいいのでは。）
%

We consider the time evolution from the initial distribution $\po$ to the final distribution $\pf$ over the finite time $\tau$.
These distributions and time are fixed throughout the optimizations discussed in the following.
The potential $V$ and the nonconservative force $\bsF$ are manipulated to drive the evolution $\po\rightarrow\pf$.
%If $\po\rightarrow\pf$ is implemented in finite time $\tau$, the total system inevitably dissipates finite amount of energy, i.e., $\sigma_{XY}>0$.
It is known that the finite-time evolution involves a finite amount of energetic dissipation, which can be minimized over the driving protocols $V(t)$, $\bsF(t)$ \cite{Aurell2011Wasserstein,Aurell2012refined}:
\begin{align}
    \label{min sigma_XY}
    \sigma_{XY}^*:=\min_{\substack{\{V,F\}_{0\leq t\leq \tau}\\ {\rm s.t.\ }\po\rightarrow\pf}}{\sigma}_{XY}
    =\frac{\calW(\po,\pf)^2}{\mu T\tau}.
\end{align}
Here, $\calW(p,q)$ represents the ($L^2$-) Wasserstein distance between the probability distributions $p$ and $q$ (with finite second moments), defined as \cite{villani2009}
\begin{align}
    \label{Wasserstein distance}
    \calW(p,q)^2:=\min_{\calT{\ \rm s.t.\ }p\rightarrow q}
    \int d\bsr\parallel \calT(\bsr)-\bsr\parallel^2 p(\bsr),
\end{align}
where $\calT$ denotes transport maps from $p$ to $q$, that is, $p(\bsr)=|\det{\left(\partial\calT(\bsr)/\partial \bsr\right)}|q(\calT(\bsr))$ holds.
The optimal protocol for achieving \eref{min sigma_XY} is constructed from the optimal transport map of \eref{Wasserstein distance} \cite{Benamou-Brenier2000,Aurell2012refined}.
Under this protocol, the probability distribution evolves along the geodesic of the Wasserstein metric between $\po$ and $\pf$.
The finite-time thermodynamic bound \eqref{min sigma_XY} has been extensively studied for both one-dimensional and Gaussian cases \cite{Dechant-Sakurai2019,Chen-2020multi-Gaussian,Nakazato-Ito2021,Proesmans2020finiteLandauer,proesmans2023precision,abiuso2022Gaussian-multi}.
%Wassersteinの活用(メモリ消去)\cite{Proesmans2020finiteLandauer}。

%The mean local velocity is related to the unique optimal transport map $\calT^{*,1/2}$ of \eref{Wasserstein distance} as $\partial_t\calT(\bsr,t)=\bsv(\calT(\bsr,t),t)$, where $\calT(\bsr,t)$ is called the Lagrange map and is the interpolation of the initial and final points of the transport map, that is, $\calT(\bsr,t)=(1-t/\tau)\bsr+(t/\tau)\calT^{*,1/2}(\bsr)$ \cite{Aurell2012refined}.
%and the initial condition $\calT(\bsr,0)=\bsr$.

%%%%%%%%%%%%%%%%%%%%%%%%%%%%%%%%%%
\textit{Main results.---}
We first overview the general structure of our optimization problem considered in this Letter.
We start with the tradeoff relation between the partial EPs of subsystems $X$ and $Y$ in finite time.
Figure \ref{fig: tradeoff} illustrates the tradeoff relation (the blue curve) as the boundary of the realizable values of the partial EPs $(\sigma_X, \sigma_Y)$ (the shaded region ).
We note that the blue curve is regarded as the Pareto front in the terminology of optimization problems.
%multi-objective optimization
The optimal values of $\sigma_X, \sigma_Y$ are denoted as $\sigma_X^*, \sigma_Y^*$, respectively.
One can further optimize the EP of $X$ (resp. $Y$) under the condition of $\sigma_Y = \sigma_Y^*$ (resp. $\sigma_X=\sigma_X^*$), which is written as $\sigma_X^* |_{\sigma_Y = \sigma_Y^*}$ (resp. $\sigma_Y^* |_{\sigma_X = \sigma_X^*}$) as indicated by the blue (resp. orange) square in \fref{fig: tradeoff}.
In general, $\sigma_X^* |_{\sigma_Y = \sigma_Y^*}$ is larger than $\sigma_X^*$  as a consequence of the tradeoff relation.
We note that the global optimal value of  \eref{min sigma_XY} corresponds to the gray square in \fref{fig: tradeoff}.

Our first main result is a general formula that connects the minimum partial EP and the Wasserstein distance (see Supplemental Material \cite{SM} for the proof):
\begin{align}
    \label{min sigma_Y general}
    \sigma_Y^*:=\min_{\substack{\{V,F\}_{0\leq t\leq \tau}\\{\rm s.t.\ }\po\rightarrow\pf}}
    {\sigma}_{Y}
    =\frac{\calW(\pyo,\pyf)^2}{\mu T\tau},
\end{align}
where $p_{Y}^{o/f}$ denote the marginal distributions of $\pt^{o/f}$ for the subsystem $Y$.
The optimal protocol is constructed from the transport map in the form of $\calT(\bsr)=[\calT_{X|Y}(x;y),\calT_Y^{*}(y)]^{\rm T}$.
Here, the optimal map for $Y$, denoted as $\calT_Y^{*}$, implements nothing but the one-dimensional optimal transport for $Y$.
On the other hand, $T_{X|Y}$ is a transport map from $p_{X|Y}^o(\cdot|y)$ to $p_{X|Y}^f(\cdot| T_Y^*(y))$.
Here, $p_{X|Y}^{o}(\cdot|y)$ and $p_{X|Y}^{f}(\cdot|\calT_Y^{*}(y))$ represent the conditional distributions of $X$ given the states $y$ and $\calT_Y^{*}(y)$ of $Y$, respectively.
%On the other hand, $\calT_{X|Y}$ transports $X$ under the condition of $Y$ being transported from $y$ to $\calT^*(y)$ such that $\calT$ reproduces the overall time evolution $\po\rightarrow\pf$.
%We note that \eref{min sigma_Y general} cannot be achieved by any $\calT_X(\bsr)$ that has $Y$ dependence \cite{SM}.
In general, there are multiple such transport maps $\calT_{X|Y}$, while we will discuss one of them later (\eref{Knothe–Rosenblatt map}).
%The protocol to achieve \eref{min sigma_Y general} can be constructed from this map.
In contrast, $\calT_Y^{*}$ is unique and constructed as the monotone map $\calT_Y^{*}(y)={\Gamma_Y^f}^{-1}(\Gamma_Y^o(y))$ \cite{santambrogio2015OT}, where $\Gamma_Y^{o/f}(y):=\int^y_{-\infty}dy'p_{Y}^{o/f}(y')$ represent the cumulative distribution functions and ${\Gamma_Y^{o/f}}^{-1}(s)$ are their inverse functions.
%${\Gamma_Y^{o/f}}^{-1}(s):=\inf_y\{\Gamma_Y^{o/f}(y)>s\}$
We note that it was pointed out in \ccite{Nakazato-Ito2021} that $\sigma_Y$ is lower-bounded by the right-hand side of \eref{min sigma_Y general}; we here found that the equality is indeed achievable as \eref{min sigma_Y general} and identified its optimal protocol.

%\eref{min sigma_Y general} says that this bound is achievable for Gaussian distributions.

We next consider the situation where the EP of $X$ is minimized after minimizing the EP of $Y$, which corresponds to the blue square in \fref{fig: tradeoff}.
%The EP of $Y$ can be minimized in the same manner as \eref{min sigma_Y general}.やはり最初から$Y$で議論しておくべきだったのでは。
Our second main result is the minimum  EP of $X$ under the condition of minimum EP of $Y$, which is given by (see Supplemental Material for the proof):
\begin{align}
    \label{min sigma_X|Y general}
    &\sigma_X^*|_{\sigma_Y=\sigma_Y^*}:=
    \min_{\substack{\{V,F\}_{0\leq t\leq \tau}{\ \rm s.t.\ }\po\rightarrow\pf\\ \sigma_Y=\sigma_Y^*}}{\sigma}_{X}\\
    &=\frac{1}{\mu T\tau}\int dy \pyo(y)\calW\left(p^o_{X|Y}(\cdot|y), p^f_{X|Y}(\cdot|\calT_Y^{*}(y))\right)^2.\nonumber
\end{align}
%Here, $p_{X|Y}^{o/f}(\cdot|y)$ represent the conditional distributions of $X$ under given $Y$'s state $y$, and $\calW\left(p^o_{X|Y}(\cdot|y), p^f_{X|Y}(\cdot|\calT_Y^{*}(y))\right)$ represents the Wasserstein distance between them.
%Here, $p_{X|Y}^{o}(\cdot|y)$ and $p_{X|Y}^{f}(\cdot|\calT_Y^{*}(y))$ represent the conditional distributions of $X$ given the states $y$ and $\calT_Y^{*}(y)$ of $Y$, respectively.
The term $\calW\left(p^o_{X|Y}(\cdot|y), p^f_{X|Y}(\cdot|\calT_Y^{*}(y))\right)$ represents the Wasserstein distance between the conditional distributions.
The optimal transport map is expressed as $\calT^{*,1}(\bsr)=[\calT_{X|Y}^{*}(x;y),\calT_Y^{*}(y)]^{\rm T}$, where the transport map for $X$ is given by the following monotone map:
\begin{align}
    \label{Knothe–Rosenblatt map}
    \calT^{*}_{X|Y}(x;y)={\Gamma_{X|Y}^f}^{-1}\left(\Gamma_{X|Y}^o(x;y);\calT_Y^{*}(y)\right).
\end{align}
Here, $\Gamma_{X|Y}^{o/f}(x;y):=\int^x_{-\infty}dx'p_{X|Y}^{o/f}(x'|y)$ denote the cumulative distribution functions of the conditional distributions and ${\Gamma_{X|Y}^{o/f}}^{-1}(s;y)$ are their inverse functions with respect to $x$.
%${\Gamma_{X|Y}^{o/f}}^{-1}(s;y):=\inf_x\{\Gamma_{X|Y}^{o/f}(x;y)>s\}$
This transport map $\calT^{*,1}$ is referred to as the Knothe–Rosenblatt map in mathematics \cite{villani2009}.
%The proofs of \esref{min sigma_Y general}\eqref{min sigma_X|Y general} are provided in Supplemental Material \cite{SM}.

The above results also hold by exchanging $X$ and $Y$.
We  note that $\sigma_Y^*$ and $\sigma_X^*|_{\sigma_Y=\sigma_Y^*}$, as well as $\sigma_{XY}^*$, are inversely proportional to the time interval $\tau$, which is a universal characteristic of the thermodynamic speed limit \cite{Shiraishi-Funo-Saito2018speed,Ito-Dechant2020info,Lee-Park_highly2022}.

%先に$\sigma_Y^*$で議論しておくべきと思ったのですが、しかし$\sigma_X^*$を議論しておいて$\sigma_X^*|_{\sigma_Y=\sigma_Y^*}$との大小関係を議論するのもアリなのかもしれません。ちなみにこの大小関係は一般に示せるのでしたっけ？
%$\sigma_X^*$のほうが小さいor一致する場合しかないことが言えます。サプリのPareto frontの一般論のところで議論します。

We describe the optimal protocol ($V^{*,1},\bsF^{*,1}$) to achieve \eref{min sigma_X|Y general} for general distributions.
Similar to the case of minimizing the total EP \cite{Aurell2012refined}, the optimal protocol is constructed as
\begin{align}
    \label{optimal protocol}
    &\mu\left(-\nabla V^{*,1}(\bsr,t)+\bsF^{*,1}(\bsr,t)-T\nabla \ln\pt(\bsr,t)\right)\nonumber\\
    &=\frac{1}{t}(\bsr-\bsr^o)_{\calT^{*,1}(\bsr^o,t)=\bsr},
\end{align}
where $\bsr^o$ represents the initial point mapped to $\bsr$ by the optimal Lagrange map $\calT^{*,1}(\bsr,t):=(1-t/\tau)\bsr+(t/\tau)\calT^{*,1}(\bsr)$.
Thus, $-\nabla V^{*,1}$ consists of the geodesic term $T\nabla \ln\pt(\bsr,t)$ and the gradient part of the right-hand side of \eref{optimal protocol}, while $\bsF^{*,1}$ comprises the nongradient part of the right-hand side.
The right-hand side of \eref{optimal protocol} functions as the counteradiabatic term ($\propto \tau^{-1}$) which drives the system along its trajectory \cite{Zhong-2024LR-OT}.
It is worth noting that the nonconservative force is required in the optimal protocol to achieve \eref{min sigma_X|Y general} (except in the uncorrelated case), which differs from the case of \eref{min sigma_XY} \cite{Aurell2011Wasserstein,Aurell2012refined}.
This is because $x^o,y^o$ in \eref{optimal protocol} become distinct gradient functions, each originating from the different maps: $\calT_{X|Y}^*,\calT_Y^*$.
%$M$とは違って$D_w^{-2}$はartificial.

\textit{Gaussian case.---}
To obtain more explicit formulas for \esref{min sigma_Y general}\eqref{min sigma_X|Y general}, we consider special cases where the initial and final distributions are Gaussian, written as $\calN(\bsm^{o/f},\Sigma^{o/f})$ with means $\bsm^{o/f}$ and covariance matrices $\Sigma^{o/f}$.
Let $\rho_{o/f}:=\Sigma_{XY}^{o/f}/\sqrt{\Sigma_{XX}^{o/f}\Sigma_{YY}^{o/f}}$ be the (Pearson) correlation coefficients.
Then, we can analytically calculate \eref{min sigma_Y general} and \eref{min sigma_X|Y general}, resulting in
\begin{align}
    \label{min sigma_Y Gaussian}
    \sigma_Y^*=\frac{(m^o_Y-m^f_Y)^2+\left(\sqrt{\syy}-\sqrt{\tyy}\right)^2}{\mu T\tau},
\end{align}
\begin{align}
    \label{min sigma_X|Y Gaussian}
    \sigma_X^*|_{\sigma_Y=\sigma_Y^*}
    =&\sigma_X^*+\frac{2\sqrt{\sxx\txx}}{\mu T \tau}\\
    &\cdot
    \left(1-\rho_o\rho_f-\sqrt{(1-{\rho_o}^{2})(1-{\rho_f}^{2})}\right)\nonumber.
\end{align}
See Supplemental Material for details.
%Thus, the effect of minimizing $\sigma_Y$ is captured by the correlation.
Applying the inequality $(a+b)/2\geq \sqrt{ab}$ for $a,b>0$, \eref{min sigma_X|Y Gaussian} is lower-bounded as
\begin{align}
    \label{evaluation of min sigma_X|Y Gaussian}
    \sigma_X^*|_{\sigma_Y=\sigma_Y^*}-\sigma_X^*
    \geq \frac{\sqrt{\sxx\txx}}{\mu T \tau}(\rho_o-\rho_f)^2\geq0.
\end{align}
%This shows that the square of the correlation change characterizes the cost of preferring one subsystem to the other subsystem.
Both the first and second inequalities in \eref{evaluation of min sigma_X|Y Gaussian} are achieved if and only if $\rho_o=\rho_f$.
In that case, the tradeoff relation (the blue curve in \fref{fig: tradeoff}) collapses to a single point, and thus there is a unique optimum shared by both subsystems.
%Therefore, there exists the tradeoff relation between $\sigma_X$ and $\sigma_Y$ if and only if the correlation is enhanced or reduced.
That is, when the correlation remains unchanged during the dynamics, $\sigma_X$ and $\sigma_Y$ can be simultaneously optimized.
%and the Pareto front consists of the single point.

\begin{figure}
    \centering
    \includegraphics[width=0.98\linewidth]{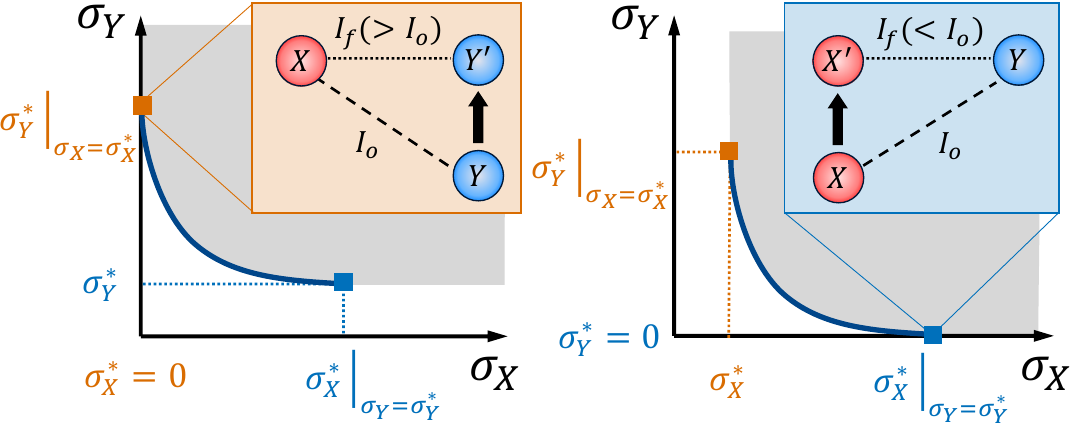}
    \caption{\label{fig: tradeoff_MF}
    Schematics for optimal measurement and feedback, which are realized at the endpoints of the Pareto front.
    At the orange (blue) square in the left (right) figure, the optimal measurement (feedback) is obtained by optimizing the EP of $Y$ ($X$) under the condition of $\sigma_X^*=0$ ($\sigma_Y^*=0$).
    The marginal distribution of $X$ ($Y$) is fixed in the left (right) figure, representing an ideal function of measurement (feedback).
    }
\end{figure}

\textit{Application to Maxwell's demons.---}
We apply our general results to Maxwell's demon setups, where $X$ is the engine and $Y$ is the demon's memory.
These two subsystems exchange information, which is quantified by the mutual information difference. 
In particular, we consider measurement and feedback processes,  illustrated in \fref{fig: tradeoff_MF}.

In the measurement process, $X$ does not evolve but $Y$ evolves depending on the state of $X$, so that they create correlation described by the mutual information difference $\Delta I=I_f-I_o>0$ (\fref{fig: tradeoff_MF} left).
%結局、測定の前後もフィードバックの前後も、同じ添え字$I_{o/f}$を使うことにしたのですね。
During this process, the marginal distributions of $X$ are fixed, and thus $\sigma_X^* = 0$ holds from \eref{min sigma_Y general} (by exchanging $X$ and $Y$).
The corresponding optimal transport map for $X$ is given by $\calT_X^*(x)=x$.
The minimum energy cost for measurement is then determined through \eref{min sigma_Y general}:
\begin{align}
    \label{work for measurement}
    \frac{W-\Delta \calF_Y}{T} - \Delta I = \sigma_Y \geq \sigma_Y^*|_{\sigma_X=\sigma_X^*},
\end{align}
where $W$ represents the work applied to $Y$ and $\calF_Y$ represents its nonequilibrium free energy \cite{Esposito2011Landauer,Parrondo-Horowitz-Sagawa2015}.
In $\tau \to \infty$, \eref{work for measurement} reduces to the bound obtained in \scite{Sagawa-Ueda2009erasure,Sagawa-Ueda2012-FT}.

We next consider the feedback process, where $Y$ functions as the memory, and the engine $X$ receives feedback based on $Y$'s record  (\fref{fig: tradeoff_MF} right). %the mutual information is consumed while keeping $Y$ fixed.
To this end, we consider the initial and final distributions such that the marginal distribution for $Y$ remains unchanged ($\pyo=\pyf$) and the mutual information is consumed through the feedback, i.e., $\Delta I=I_f-I_o<0$.
In this case, $\sigma_Y^* = 0$ holds  from \eref{min sigma_Y general} and the corresponding optimal transport of $Y$ is given by $\calT_Y^*(y)=y$.
The optimal extractable work via feedback is determined through \eref{min sigma_Y general}:
\begin{align}
    \label{extractable work by demon}
    \frac{-W_{\rm ext}-\Delta \calF_X}{T} - \Delta I = \sigma_X \geq \sigma_X^*|_{\sigma_Y=\sigma_Y^*},
\end{align}
where $W_{\rm ext}$ represents the work extracted from $X$ and $\calF_X$ represents its nonequilibrium free energy.
In $\tau \to \infty$,  \eref{work for measurement} reduces to the bound obtained in \scite{Sagawa-Ueda2008quantumfeedback,Sagawa-Ueda2012-FT}.
%with $\sigma_X^*|_{\sigma_Y=\sigma_Y^*}=\frac{1}{\mu T\tau}\int dy \pyo(y)\calW\left(p^o_{X|y}, p^f_{X|y}\right)^2$, where $p^{o/f}_{X|y}$ denotes the conditional distributions of $X$ given $Y$'s state $y$.
We note that, while a bound of the form \eref{extractable work by demon} has also been obtained in \ccite{Taghvaei-2022-demon}, 
 the optimal protocol to achieve \eref{extractable work by demon} and explicit models to realize optimal finite-time demons were unexplored in literature.

We next show that when the operation time $\tau$ is too short for the feedback process, Maxwell's demon cannot be realized.
Due to the time-dependence of $\sigma_X^*|_{\sigma_Y=\sigma_Y^*}(\propto\tau^{-1})$, there exists a characteristic time $\tau_{\textrm{demon}}(<\infty)$ at which $\sigma_X^*|_{\sigma_Y=\sigma_Y^*}=-\Delta I$ holds.
If the feedback is implemented sufficiently slowly ($\tau>\tau_{\textrm{demon}}$) and optimally, the extracted work $W_{\rm ext}$ can exceed the free energy difference $-\Delta \calF_X$, thereby realizing Maxwell's demon.
In contrast, Maxwell's demon cannot be realized, regardless of optimization, for fast processes satisfying $\tau\leq\tau_{\textrm{demon}}$.
For non-optimal protocols, $\tau_{\textrm{demon}}$ can also be defined in a similar manner.
This is because the constraint $\po\rightarrow\pf$ remains satisfied by varying the operation time $\tau$ as long as the $t/\tau$ dependency remains the same.
Since $\tau_{\textrm{demon}}$ is minimized for the optimal protocol, this parameter characterizes the optimality of the given protocol.
We note that the feasibility of implementing Maxwell's demon in finite time has also been studied for discrete systems in \ccite{kamijima2024discrete}.

%Calculations of \esref{min sigma_X|Y Gaussian}\eqref{apdx: optimal map XY Gaussian} are listed in Supplemental Material \cite{SM}.

%While we assume that the mobility is isotropic, 
\begin{figure}
    \centering
    \includegraphics[width=0.98\linewidth]{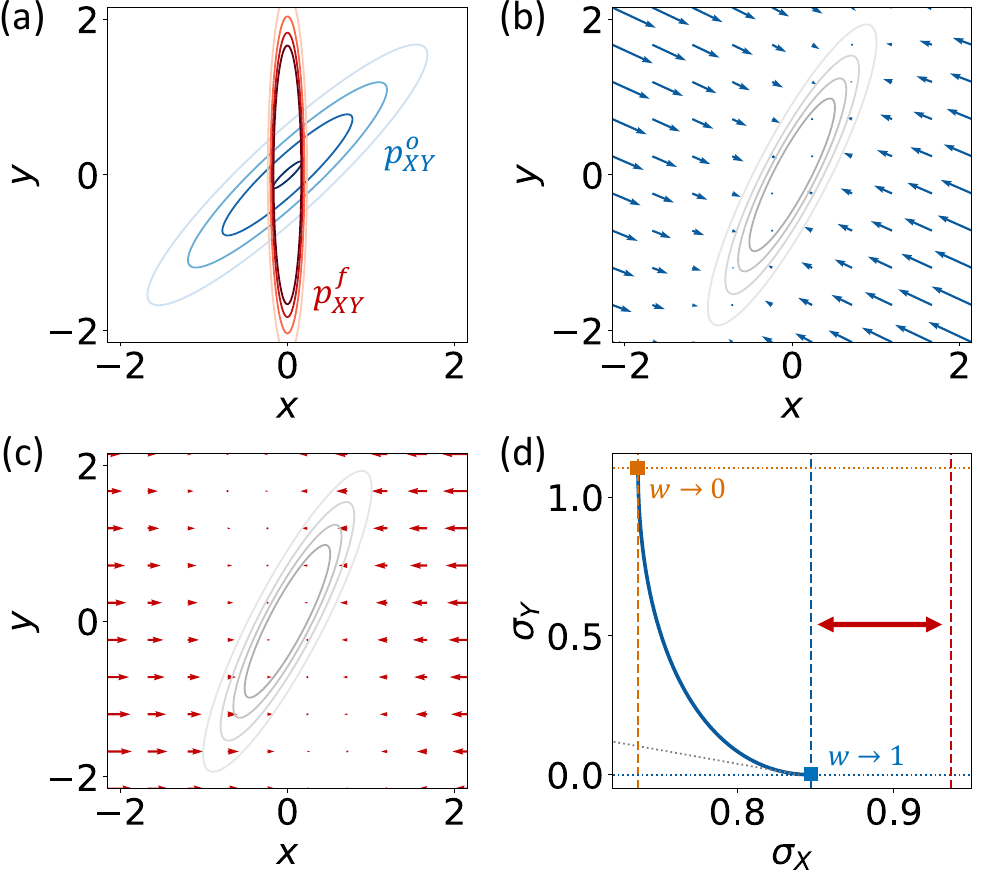}
    \caption{\label{fig: feedback_Gaussian}
    An optimal Maxwell's demon performing feedback for Gaussian distributions.
    (a) The initial and final distributions (\eref{feedback dist Gaussian}).
    (b)(c) The optimal protocol and distribution (gray contours) at $t/\tau=1/2$.
    The forces of \eref{optimal protocol} at each point are depicted as vectors, where the magnitudes of the forces are represented by the lengths of the vectors.
    (b) The geodesic term $T\nabla \ln\pt$ (scaled by $1/100$). 
    (c) The counteradiabatic term $(\bsr-\bsr^o)/\mu t$ (scaled by $1/10$).
    (d)  The optimal cost of feedback (the blue square).
    The blue curve corresponds to that in (see also \fref{fig: tradeoff}) and represents the tradeoff relation between $\sigma_X$ and $\sigma_Y$.
    The vertical dotted lines represent $\sigma_X^*$ (orange), $\sigma_X^*|_{\sigma_Y=\sigma_Y^*}$ (blue), and $-\Delta I$ (red).
    The red arrow indicates that $\sigma_X^*|_{\sigma_Y=\sigma_Y^*}<-\Delta I$, and therefore Maxwell's demon is realized in this protocol (see \eref{extractable work by demon}).
    The gray line shows the minimization of the total EP (\eref{min sigma_XY}). 
    The parameters are set to $\sxx=1.0,\txx=0.010,\syy=\tyy=l=1.0,\rho_o=0.92,\rho_f=0,\tau=1.1$.
    }
\end{figure}

\textit{Gaussian model of Maxwell's demon.---}
We demonstrate an optimal feedback process that realizes Maxwell's demon in finite time (\eref{extractable work by demon}) by a Gaussian model.
Let us first consider the initial and final distributions in Gaussian form, $\pt^{o/f}=\calN(\bsm^{o/f},\Sigma^{o/f})$, which satisfy the following conditions:
\begin{align}
    \label{feedback dist Gaussian}
    \begin{cases}
        m^o_X=m^f_X, \Sigma^o_{XX}\gg\Sigma^f_{XX}\\
        m^o_Y=m^f_Y, \Sigma^o_{YY}=\Sigma^f_{YY}\\
        \rho_o>\rho_f\geq0
    \end{cases}.
\end{align}
These distributions are illustrated in \fref{fig: feedback_Gaussian}(a).
The first condition in \eref{feedback dist Gaussian} indicates that the variance (fluctuation) of $X$ is reduced while maintaining the same average.
As a result of the second condition, the marginal distribution for $Y$ remains unchanged throughout the process.
Since the mutual information is written in terms of the correlation as $I_{o/f}=-\frac{1}{2}\ln(1-{\rho_{o/f}}^2)$ for Gaussian distributions, the last condition implies that $Y$'s information about $X$ is consumed during the process.
We note that this last condition can be replaced by $\rho_o<\rho_f\leq0$.
For simplicity, we further assume $\bsm^{o/f}=0$ and $\rho_f=0$ in what follows.

We now formulate the optimal feedback process for the distributions in \eref{feedback dist Gaussian} using \eref{min sigma_X|Y Gaussian}.
In this process, the engine $X$ receives feedback to return to $x\simeq0$ based on $Y$'s record, $y$ ($\simeq x$).
The optimal cost of feedback is calculated from \eref{min sigma_X|Y Gaussian} as
\begin{align}
    \label{optimal feedback cost}
    \sigma_X^*|_{\sigma_Y=\sigma_Y^*}
    =\sigma_X^*+\frac{2\sqrt{\sxx\txx}}{\mu T\tau}\left(1-e^{-I_o}\right).
\end{align}
This suggests that greater energetic dissipation is required when $\tau$ is reduced (faster feedback), $I_o$ is increased (greater information consumption), or $\sxx$ is increased (longer feedback displacement).
%We note that the deviation from $\sigma_X^*$ in \eref{optimal feedback cost} is proportional to the initial mutual information for small $\rho_o$.
%Y静止なら、非保存力を保存力に埋め込める

\begin{figure}[t]
    \centering
    \includegraphics[width=0.95\linewidth]{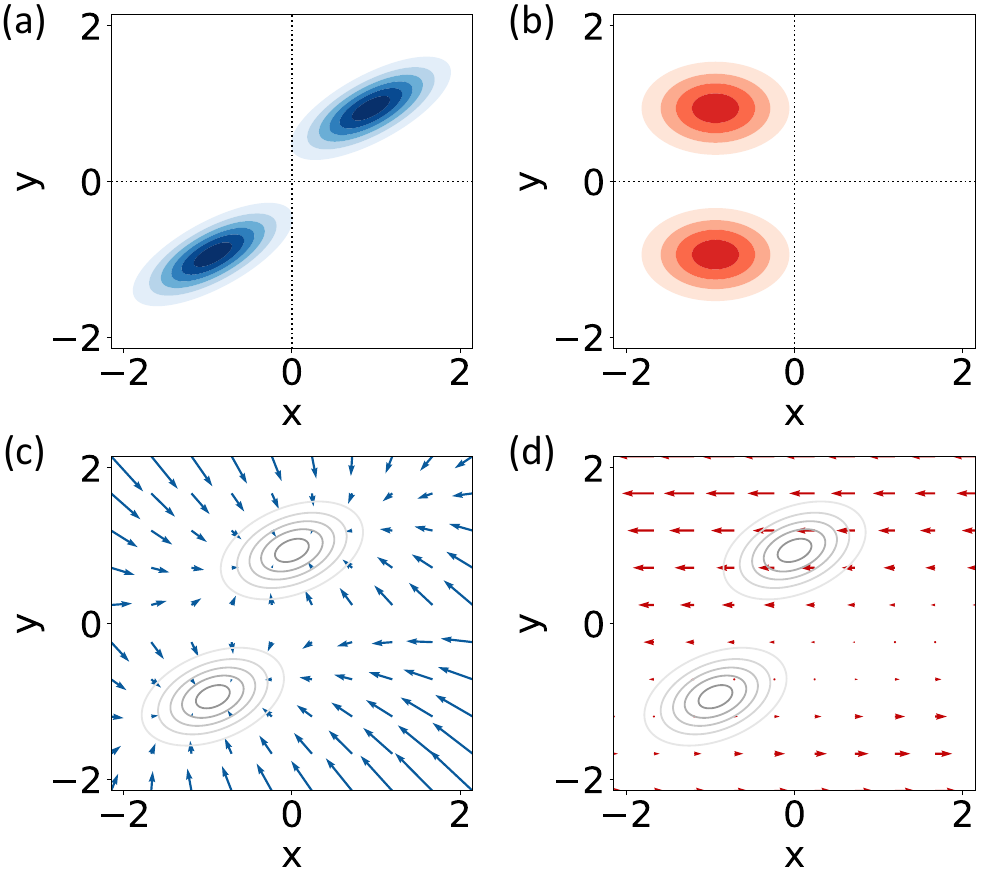}
    \caption{\label{fig: feedback_double}
    An optimal Maxwell's demon performing feedback with the double-well potentials.
    (a) The initial distribution (\eref{feedback dist double}).
    (b) The final distribution (\eref{feedback dist double}).
    (c)(d) The optimal protocol and distribution (gray contours) at $t/\tau=1/2$.
    The forces of \eref{optimal protocol} at each point are depicted as vectors, where the magnitudes of the forces are represented by the lengths of the vectors.
    (c) The geodesic term $T\nabla \ln\pt$ (scaled by $1/40$) 
    (d) The counteradiabatic term $(\bsr-\bsr^o)/\mu t$ (scaled by $1/4$).
    The parameters are set to $a=0.34,b=0.94,c=0.35,d=0.50,l=\sqrt{\textrm{Var}Y}=1.0,\tau=2.2$.
    }
\end{figure}

We explain the optimal protocol to achieve \eref{optimal feedback cost}.
For Gaussian distributions, the optimal transport map $\calT^{*,1}$ is linear (more precisely, affine).
The corresponding protocol is calculated from \eref{optimal protocol} and the forces also take the linear form.
The nonconservative force comes from the term in the right-hand side of \eref{optimal protocol} that has $y$ dependency.
Figure \ref{fig: feedback_Gaussian}(c)(d) illustrates this optimal protocol at the intermediate time $t/\tau=1/2$ (see Supplemental Material for details).
%The geodesic term $T\nabla \ln\pt$ is dominant for the given operation time $\tau$ considered in \fref{fig: feedback_Gaussian},
The counteradiabatic term $(\bsr-\bsr^o)/\mu t$ functions to suppress the fluctuation of $X$.
For the parameters in \fref{fig: feedback_Gaussian}, Maxwell's demon is realized ($\sigma_X^*|_{\sigma_Y=\sigma_Y^*}=0.85,-\Delta I =0.94$) and the time limit for the demon is calculated as
$\tau_{\textrm{demon}}=0.99 l^2/\mu T$,
where $l:=\sqrt{\Sigma_Y^{o/f}}$ denotes the standard deviation of $Y$. 
We set $\mu=T=1$ in our numerical calculations.

%As the time passes, the system has to be confined strongly in the $Y$ direction (see \fref{fig: feedback_Gaussian}(a)), leading to the large value of the $XX$ components.

We note that optimization of finite-time feedback for Gaussian distributions has also been studied in \ccite{abreu-Seifert2011feedback} in a slightly different setup.
Their optimal work can be expressed as a special case of  \eref{min sigma_X|Y general}, while \eref{min sigma_X|Y general} applies to general non-Gaussian distributions.

\textit{Non-Gaussian demon with double-well potentials.---}
We next demonstrate a non-Gaussian optimal feedback process by Maxwell's demon using double-well potentials.
We consider the following initial and final distributions (see also \fref{fig: feedback_double}(a)(b)):
\begin{align}
    \label{feedback dist double}
    \begin{cases}
        p^{o/f}(y)\propto \frac{1}{2}e^{-\frac{1}{2a^2}(y-b)^2}+\frac{1}{2}e^{-\frac{1}{2a^2}(y+b)^2}\\
        p^{o}_{X|Y}(x|y)\propto e^{-\frac{1}{2c^2}(x-y)^2},p^{f}_{X|Y}(x|y)\propto e^{-\frac{1}{2d^2}(x+b)^2},
    \end{cases}
\end{align}
where $a,b,c,d$ are the fixed parameters.
The memory $Y$ is initially located around $y=\pm b$ and is highly correlated to the engine $X$.
At the end of this process, the correlation disappears and $X$ is mainly located around $x=-b$.
Since the marginal distribution for $Y$ is identical for $t=0$ and $t=\tau$, the minimization \eqref{min sigma_X|Y general} fixes $Y$ and implements the optimal feedback process as in the Gaussian case.
This feedback moves $X$ to $x\simeq -b$ if $x\simeq b$ initially and keeps $x\simeq -b$ if $x\simeq -b$ initially.

The optimal feedback cost is given by \eref{extractable work by demon} and is calculated as $\sigma_X^*|_{\sigma_Y=\sigma_Y^*}=[a^2+2b^2+(c-d)^2]/\mu T\tau$.
The optimal map is calculated as $\calT_{X|Y}^{*}(\bsr)=\frac{d}{c}(x-y)-b,\calT_Y^*(y)=y$.
The optimal protocol is obtained from \eref{optimal protocol}, as shown in \fref{fig: feedback_double}(c)(d).
The counteradiabatic term $(\bsr-\bsr^o)/\mu t$ provides feedback to $X$ to move toward $x\simeq -b$ when $y\simeq b$.
%The forces are also linear due to the form of \eref{feedback dist double}.
%混合Gaussだとnonlinear, step関数で分けるとlinear
For the parameters in \fref{fig: feedback_double}, Maxwell's demon is realized ($\sigma_X^*|_{\sigma_Y=\sigma_Y^*}=0.87,-\Delta I =0.95$) and the time limit for the demon is calculated as $\tau_{\textrm{demon}}=2.0 l^2/\mu T$.
The time $\tau_{\textrm{demon}}$ is nearly twice the Gaussian case, while $-\Delta I$ remains nearly the same.
We note that $\tau_{\textrm{demon}}$ can be improved by optimizing the parameters $a,b,c,d$.
%プロトコルの説明.
%数値的にはHelmholtzの分解定理.
%EPの計算.

%%%%%%%%%%%%%%%%%%%%%%%%%%%%%%%%%%

%%%%%%%%%%%%%%%%%%%%%%%%%%%%%%%%%%
\textit{Conclusion and perspectives.---}
For bipartite overdamped Langevin systems, we derived the fundamental thermodynamic bounds for the partial EPs in finite time (\esref{min sigma_Y general}\eqref{min sigma_X|Y general}).
By adopting optimal transport theory, we showed that the bounds are achievable for any time interval and identified the optimal protocol.
While our general results apply to non-Gaussian cases (\eref{min sigma_X|Y general}), the concise formulas are derived for the Gaussian cases (\eref{min sigma_X|Y Gaussian}).
Based on our theory, we considered measurement and feedback by Maxwell's demon and demonstrated optimal feedback processes by the demons for both Gaussian and non-Gaussian models (\fsref{fig: feedback_Gaussian}, \ref{fig: feedback_double}).

We here make some remarks.
First, the optimal EPs (\esref{min sigma_Y general}\eqref{min sigma_X|Y general}) cannot be derived from the continuous limit of the discrete Markov jump case \cite{kamijima2024discrete}. 
This is because such limit yields a lower bound of the EP in terms of $L^1$, instead of $L^2$, Wasserstein distance \cite{dechant2022minimum,Van-Saito-unification2023}, while the bound given by the $L^1$-Wasserstein distance is not achievable in general. 
Therefore, it is crucial to treat Langevin cases separately from Markov jumps cases.
%Second, while we mainly focused on the feedback processes in our examples, the optimal measurement can be argued in the same manner by applying our general results (\esref{min sigma_X|Y general}\eqref{min sigma_X|Y Gaussian}).
%Nonetheless, it is an intriguing direction to explore the tradeoff relation between the measurement cost and feedback cost \cite{kamijima2024discrete} for non-Gaussian distributions.
Second, the nonconservative force needs to be manipulated time-dependently in the optimal feedback protocol (\eref{optimal protocol}), which would be challenging even with state-of-the-art experimental techniques of optical tweezers \cite{sukhov-2017nonconservative-optical}.
%光トラップに非保存力が存在するという文献は多数あるが、engineerできるかは書かれていない
It is an interesting future direction to implement various optimal thermodynamic processes in real experiments, ranging from colloidal particles to biomolecules \cite{Saha-2022Bayesian-exp,tang-2024SC-szilard}.
%3次元以上（可能だが面倒）.
%Gaussianのトレードオフのあるなしは相関係数で簡単に判別できるが一般則は不明.
%また、始分布・終分布をGaussianに限らないクラスの中で最適化して(Nagase引用)、同じ$\Delta I$でより短い$\tau_{\textrm{demon}}$を求めるということもできる。

\begin{acknowledgments}
\textit{Acknowledgement.---}
We thank Kaito Tojo for helpful feedback on the manuscript.
This work is supported by JST ERATO Grant Number  JPMJER2302, Japan.
T.K. is supported by World-leading Innovative Graduate Study Program for Materials Research, Information, and Technology (MERIT-WINGS) of the University of Tokyo.
T.K. is also supported by JSPS KAKENHI Grant No.JP24KJ0611.
A.T. acknowledges support from JSPS KAKENHI (Grant Nos. 19H05599. and 19K03494).
K.F. acknowledges support from JSPS KAKENHI (Grant Nos. JP23K13036 and JP24H00831).
T.S. is supported by JSPS KAKENHI Grant No. JP19H05796 and JST CREST Grant No. JPMJCR20C1. T.S. is also supported by Institute of AI and Beyond of the University of Tokyo.
\end{acknowledgments}

%\bibliography{ref}
%

\end{document}

%\onecolumngrid

\setcounter{section}{0}
\setcounter{equation}{0}
\setcounter{figure}{0}
\setcounter{table}{0}
\setcounter{page}{1}
\renewcommand{\thesection}{S\arabic{section}}
\renewcommand{\theequation}{S\arabic{equation}}
\renewcommand{\thefigure}{S\arabic{figure}}
\renewcommand{\thetable}{S\arabic{table}}
\renewcommand{\bibnumfmt}[1]{[S#1]}
\begin{center}
	\Large
	\textbf{Supplemental Material for ``Finite-time optimal Maxwell's Demons''}
\end{center}

\section{Review of Stochastic Thermodynamics}
\label{apdx:}
%We here review briefly stochastic and information thermodynamics.
First, we briefly review stochastic thermodynamics of overdamped Langevin systems (Eqs.~(\blue{1})(\blue{2})).
The rates of the absorbed heat and extracted work are defined as \cite{Seifert2012}
\begin{align}
    &\dot{Q}:=-\int d\bsr (-\nabla V(\bsr,t)+\bsF(\bsr,t))^{\rm T} \bsv(\bsr,t)\pt(\bsr,t),\\
    &\dot{W}_{\rm ext}:=\int d\bsr (-\partial_t V(\bsr,t)-\bsF(\bsr,t)^{\rm T}\bsv(\bsr,t)) \pt(\bsr,t).
\end{align}
The average energy of the system is given by $E :=\int d\bsr V(\bsr,t)\pt(\bsr,t)$, leading to the first law of thermodynamics  $d_tE=-\dot{W}_{\rm ext}+\dot{Q}$.
By using the Shannon entropy of the total system
\begin{align}
    S_{XY}:=- \int d\bsr \pt(\bsr,t)\ln \pt(\bsr,t),
\end{align}
the entropy production rate of the total system [Eq.~(\blue{3}) of the main text] is given by
\begin{align}
    &\dot{\sigma}_{XY}=d_t{S}_{XY}-\frac{\dot{Q}}{T}.
\end{align}
By introducing the nonequilibrium free energy
\begin{align}
\calF_{XY}:=E-TS_{XY},
\end{align}
the entropy production rate is rewritten as
\begin{align}
    &\dot{\sigma}_{XY}=\frac{-\dot{W}_{\rm ext}-d_t{\calF}_{XY}}{T}.
\end{align}

In bipartite systems, the mutual information between subsystems is defined as \cite{cover1999elements}
\begin{align}
    I:= \int d\bsr \pt(\bsr,t)\ln \frac{\pt(\bsr,t)}{\pxt(x,t)\pyt(y,t)}.
\end{align}
The change of $I$ can be decomposed into the contributions of the subsystems \cite{allahverdyan-2009infoflow,infoflow-Holowitz-2014}:
\begin{align}
    &d_tI=\dot{I}_X+\dot{I}_Y,\\
    &\dot{I}_X:=\int d\bsr \pt(\bsr,t)v_X(\bsr,t)\partial_x\ln \frac{\pt(\bsr,t)}{\pxt(x,t)\pyt(y,t)}, \\
    &\dot{I}_Y:=\int d\bsr \pt(\bsr,t)v_Y(\bsr,t)\partial_y\ln \frac{\pt(\bsr,t)}{\pxt(x,t)\pyt(y,t)},
\end{align}
where $\dot{I}_X$ ($\dot{I}_Y$) is called the information flow and provides the information gain or loss induced by the change of $X$ ($Y$).
The entropy production rate of the subsystem $X$ (Eq.~(\blue{4})) is written as
\begin{align}
    \label{apdx: sigma_X}
    \dot{\sigma}_{X}=d_t{S}_{X}-\frac{\dot{Q}_X}{T}-\dot{I}_X,
\end{align}
where $S_{X}:=- \int dx \pxt(x,t)\ln \pxt(x,t)$ is the Shannon entropy of $X$ and $\dot{Q}_X:=-\int d\bsr (-\partial_x V(\bsr,t)+F_X(\bsr,t))v_X(\bsr,t)\pt(\bsr,t)$ is the heat absorbed by $X$.
%The terms in \eref{apdx: sigma_X} are attributed to the subsystem $X$.
%However, this attribution gets unclear if we use the work instead of the heat in the expression.
In the special case that the other subsystem $Y$ does not evolve as in the feedback process, the EP of $X$ is expressed using the work:
\begin{align}
    \dot{\sigma}_X=\dot{\sigma}_{XY}=
    \frac{-\dot{W}_{\rm ext}-d_t\calF_{X}}{T}-d_tI,
\end{align}
where $\calF_{X}:=E-TS_{X}$ is the nonequilibrium energy of the subsystem $X$.
The same definitions and formulas also apply to the subsystem $Y$ by exchanging $X$ and $Y$.
%(denoted as $\calF$ in \esref{quasistatic demon}\eqref{finite demon})

%\section{Consistensy with \ccite{abreu-Seifert2011feedback}}

%%%%%%%%%%%%%%%%%%%%%%%%%%%%%%%%%%
\section{The Optimal Feedback Protocol}
In this section, we explain the optimal map and the optimal protocol for the finite-time feedback in Figs.~\blue{3},\blue{4}.

\subsection{The Gaussian Case}

\subsubsection{The optimal map}
We here consider general Gaussian distributions $\pt^{o/f}=\calN(\bsm^{o/f},\Sigma^{o/f})$ for illustration, while Fig.~\blue{3} assumes a specific form of distributions expressed in Eq.~(\blue{16}).
Although the optimal maps for $X$ and $Y$ can be calculated using the cumulative distribution functions, as stated in the main text, here we leverage the fact that if a one-dimensional transport map has a non-decreasing linear form then it is optimal \cite{villani2003topics}.
We first explicitly present the optimal transport map for $Y$, assumed to be $\calT_Y^*(y):=a_Y y+b_Y$ ($a_Y\geq0$).
Since the marginal distribution for $Y$ is transported by this map as
\begin{align}
    \pyf(y)=\int dy^o\delta(y-(a_Y y^o+b_Y))\pyo(y^o)
    =\frac{1}{a_Y}\pyo\left(\frac{y-b_Y}{a_Y}\right),
\end{align}
the constants $a_Y$ and $b_Y$ are calculated as $a_Y=\sqrt{{\tyy}/{\syy}}$ and $b_Y=m_Y^f-a_Y m_Y^o$, respectively.
Thus, the optimal map for $Y$ is given by
\begin{align}
    \label{apdx: optimal map Y Gaussian}
    \calT_Y^*(y)=\sqrt{\frac{\tyy}{\syy}}(y-m_Y^o)+m_Y^f.
\end{align}
%As observed from this, $\calT_Y^*$ normalizes $Y$ from $\pyo$ and biases it toward $\pyf$.

We next construct the $X$ component of the transport map, assumed to be $\calT_{X|Y}^*(x;y):=a_X x+b_X$ ($a_X\geq0$).
The conditional distributions of $X$ given $Y=y$ are
\begin{align}
    \label{apdx: conditional distribution Gaussian}
    p_{X|Y}^{o/f}(x|y)=\sqrt{\frac{\styy}{2\pi\det{\Sigma^{o/f}}}}
    \exp\left[-\frac{\styy}{2\det{\Sigma^{o/f}}}\left(x-m_X^{o/f}-\frac{\stxy}{\styy}(y-m_Y^{o/f})\right)^2\right].
\end{align}
Since $p_{X|Y}^o(\cdot|y)$ is transported to $p_{X|Y}^f(\cdot|T_Y^*(y))$ by the map $\calT_{X|Y}^*$,
\begin{align}
    p_{X|Y}^f(x|\calT_Y^*(y))=
    \int dx^o\delta\left(x-(a_X x^o+b_X)\right)p_{X|Y}^o(x^o|y)
    =\frac{1}{a_X}p_{X|Y}^o\left(\frac{x-b_X}{a_X}\Big|y\right)
\end{align}
holds.
The constants $a_X$ and $b_X$ are determined by comparison, yielding the transport map for $X$ as
\begin{align}
    \calT^{*}_{X|Y}(x;y)&=
    \sqrt{\frac{\det{\Sigma^{f}}}{\tyy}\frac{\syy}{\det{\Sigma^{o}}}}
    \left(x-m_X^{o}-\frac{\sxy}{\syy}(y-m_Y^{o})\right)
    +m_X^f+\frac{\txy}{\tyy}\left(\sqrt{\frac{\tyy}{\syy}}(y-m_Y^o)+m_Y^f-m_Y^f\right)\nonumber\\
    \label{apdx: optimal map X Gaussian}
    &=\sqrt{\frac{1-{\rho_f}^{2}}{1-{\rho_o}^{2}}\frac{\txx}{\sxx}}(x-m_X^o)
    +\left(\rho_f-\rho_o\sqrt{\frac{1-{\rho_f}^{2}}{1-{\rho_o}^{2}}}\right)\sqrt{\frac{\txx}{\syy}}(y-m_Y^o)+m_X^f.
\end{align}

To summarize, the optimal transport map to achieve Eq.~(\blue{12}) is represented as
\begin{align}
   \label{apdx: optimal map XY Gaussian}
    &\calT^{*,1}(\bsr)=G_{1}(\bsr-\bsm^o)+\bsm^f,\\
    \label{apdx: G1}
    &G_{1}=
    \begin{bmatrix}
    \sqrt{\frac{1-{\rho_f}^{2}}{1-{\rho_o}^{2}}\frac{\txx}{\sxx}}\ \ \  & \left(\rho_f-\rho_o\sqrt{\frac{1-{\rho_f}^{2}}{1-{\rho_o}^{2}}}\right)\sqrt{\frac{\txx}{\syy}} \\
    0 & \sqrt{\frac{\tyy}{\syy}}
    \end{bmatrix}.
\end{align}

\subsubsection{The optimal protocol}
We next construct the optimal protocol from the obtained optimal map (\eref{apdx: optimal map XY Gaussian}).
We set $\syy=\tyy$, $\bsm^{o/f}=0$, and $\rho_f=0$ as in the main text.
The Lagrange map is given by
$\calT^{*,1}(\bsr,t)=(1-t/\tau)\bsr+(t/\tau)\calT^{*,1}(\bsr)=\left((1-t/\tau)\bm{1}+(t/\tau)G_1\right)\bsr$, with $\bm{1}$ being the $2 \times 2$ identity matrix.
The covariance matrix at time $t$ is written as
\begin{align}
    \Sigma(t) = \left(\left(1-\frac{t}{\tau}\right)\bm{1}+\frac{t}{\tau}G_1\right)
    \Sigma^o \left(\left(1-\frac{t}{\tau}\right)\bm{1}+\frac{t}{\tau}G_1\right)^{\rm T}.
\end{align}
The boundary condition $\Sigma(\tau)=\Sigma^f$ is verified by $G_1\Sigma^oG_1^{\rm T}=\Sigma^f$.
The right-hand side of Eq.~(\blue{10}) is calculated as follows:
\begin{align}
    \bsr^o&=\left(\left(1-\frac{t}{\tau}\right)\bm{1}+\frac{t}{\tau}G_1\right)^{-1}\bsr\nonumber\\
    &=
    \begin{bmatrix}
    \frac{1}{1-\frac{t}{\tau}+\frac{t}{\tau}\sqrt{\txx/(1-{\rho_o}^2)\sxx}} 
    & \frac{\frac{t}{\tau}\rho_o\sqrt{\txx/(1-{\rho_o}^2)\syy}}{1-\frac{t}{\tau}+\frac{t}{\tau}\sqrt{\txx/(1-{\rho_o}^2)\sxx}} \\
    0 & 1
    \end{bmatrix}\bsr,\\
    \label{apdx: counteradiabatic Gauss}
    \frac{\bsr-\bsr^o}{t}
    &=
    \begin{bmatrix}
    \frac{-\tau^{-1}}{1-\frac{t}{\tau}+\frac{t}{\tau}\sqrt{\txx/(1-{\rho_o}^2)\sxx}}
    \left(\left(1-\sqrt{\txx/(1-{\rho_o}^2)\sxx}\right)x+\rho_o\sqrt{\txx/(1-{\rho_o}^2)\syy}y \right)
    \\0
    \end{bmatrix}.
\end{align} 
The nonconservative force is required to implement the term of $y$ in \eref{apdx: counteradiabatic Gauss}.
%From Eq.~(\blue{10}), the optimal forces are given by
%\begin{align}
%    &-\nabla V^{*,1}(\bsr,t)
%    =-T\Sigma(t)^{-1}\bsr-\frac{1}{\mu \tau}\frac{1-\sqrt{\txx/(1-{\rho_o}^2)\sxx}}{1-\frac{t}{\tau}+\frac{t}{\tau}\sqrt{\txx/(1-{\rho_o}^2)\sxx}}x\bse_0,\\
%    &\bsF^{*,1}(\bsr,t)=
%    -\frac{1}{\mu \tau}\frac{\rho_o\sqrt{\txx/(1-{\rho_o}^2)\syy}}{1-\frac{t}{\tau}+\frac{t}{\tau}\sqrt{\txx/(1-{\rho_o}^2)\sxx}}y\bse_0,
%\end{align}
%where $\bse_0=[1,0]^{\rm T}$.
The optimal protocol is depicted in Fig.~\blue{3}(a) and (b).

\subsection{The Double-Well Case}
\subsubsection{The optimal map}
We consider the distributions expressed in Eq.~(\blue{18}).
Since the marginal distribution of $Y$ is identical at $t=0$ and $t=\tau$, the optimal transport for $Y$ is simply $\calT_Y^*(y)=y$.
The cumulative distribution functions of $p_{X|Y}^{o/f}$ are given by 
\begin{align}
    \Gamma_{X|Y}^{o}(x;y)=\Gamma\left(\frac{1}{c}(x-y)\right),\Gamma_{X|Y}^{f}(x;y)=\Gamma\left(\frac{1}{d}(x+b)\right),
\end{align}
where $\Gamma(x):=\int^x_{-\infty}dx'\exp(-x'^2/2)/\sqrt{2\pi}$ represents the cumulative distribution function of the normal distribution.
The inverse function of $\Gamma_{X|Y}^f$ with respect to $x$ is
\begin{align}
    {\Gamma_{X|Y}^{f}}^{-1}(s;y)=d\ \Gamma^{-1}(s)-b,
\end{align}
where $\Gamma^{-1}$ denotes the inverse function of $\Gamma$.
From Eq.~(\blue{9}), the $X$ component of optimal transport map is obtained as
\begin{align}
    \label{apdx: optimal map X double}
    \calT^{*}_{X|Y}(x;y)={\Gamma_{X|Y}^f}^{-1}\left(\Gamma_{X|Y}^o(x;y);\calT_Y^{*}(y)\right)=\frac{d}{c}(x-y)-b.
\end{align}
%The EP of $X$ is calculated from Eq.~(\blue{8}).

\subsubsection{The optimal protocol}
We construct the optimal protocol from $\calT_Y^*(y)=y$ and \eref{apdx: optimal map X double}.
The probability distribution evolves to
\begin{align}
    p_{XY}(\bsr,t)=\int d\bsr^o\delta(\bsr-\calT^{*,1}(\bsr^o,t))\po(\bsr^o)
    =\frac{1}{(1-\frac{t}{\tau})+\frac{t}{\tau}\frac{d}{c}}
    p_{X|Y}^o\left(\frac{x+\frac{t}{\tau}(\frac{d}{c}y+b)}{(1-\frac{t}{\tau})+\frac{t}{\tau}\frac{d}{c}}\Big|y\right)\pyo(y)
\end{align}
at time $t$, where $\calT^{*,1}(\bsr^o,t)$ is the Lagrange map in Eq.~(\blue{10}). 
The right-hand side of Eq.~(\blue{10}) is calculated as
\begin{align}
    &\bsr=\calT^{*,1}(\bsr^o,t)=
    \begin{bmatrix}
    (1-\frac{t}{\tau})x^o+\frac{t}{\tau}\frac{d}{c}(x^o-y^o)-\frac{t}{\tau}b\\
    y^o
    \end{bmatrix},
    \\
    \label{apdx: counteradiabatic double}
    &\frac{\bsr-\bsr^o}{t}=
    \begin{bmatrix}
    \frac{-1}{\tau}\frac{(1-\frac{d}{c})x+b+\frac{d}{c}y}{(1-\frac{t}{\tau})+\frac{t}{\tau}\frac{d}{c}}\\
    0
    \end{bmatrix}.
\end{align}
The nonconservative force is required to implement the term of $y$ in \eref{apdx: counteradiabatic double}.
%The forces in the optimal protocol are then given by
%\begin{align}
%    &-\nabla V^{*,1}(\bsr,t)
%    =T\nabla\ln p_{XY}(\bsr,t)-
%    \frac{1}{\mu\tau}\frac{(1-\frac{d}{c})x+b}{(1-\frac{t}{\tau})+\frac{t}{\tau}\frac{d}{c}}\bse_0,\\
%    &\bsF^{*,1}(\bsr,t)=-
%    \frac{1}{\mu\tau}\frac{\frac{d}{c}y}{(1-\frac{t}{\tau})+\frac{t}{\tau}\frac{d}{c}}\bse_0.
%\end{align}
%Since $\nabla\ln p_{XY}(\bsr,t)$ is linear (see Eq.~(\blue{18})), both the forces have linear forms.
The optimal protocol is depicted in Fig.~\blue{4}(c) and (d).
%%%%%%%%%%%%%%%%%%%%%%%%%%%%%%%%%%

\section{Proofs of the Main Results}

In this section, we prove the main results of the main texts.
In what follows, we always assume that all functions and maps are regular enough to justify our computation.

\subsection{Proof of Eq.~(\blue{7})}
\label{subsec: proof of min sigma_X}
%In this subsection, we denote the transport map as $\calT^f$ instead of $\calT$ for convenience.
To begin with, we change the variable from the velocity field $\{\bsv\}_{0\leq t\leq\tau}$ to the Lagrange map $\{\calT\}_{0\leq t\leq\tau}$ that realizes $\po\rightarrow\pf$.
This map satisfies $\partial_t \calT(\bsr,t)=\bsv(\calT(\bsr,t),t)$, with the initial condition $\calT(\bsr,0)=\bsr$ and the final condition $\calT(\bsr,\tau)=\calT^f(\bsr)$, where $\calT^f$ (denoted as $\calT$ in the main text) is a transport map from $\po$ to $\pf$.
Then, the EP of $Y$ is evaluated as
\begin{align}
    \label{apdx: proof of min sigma_Y}
    \sigma_Y&=\frac{1}{\mu T}\int_0^\tau dt\int d\bsr v_Y(\bsr,t)^2 \pt(\bsr,t)\nonumber\\
    &=\frac{1}{\mu T}\int_0^\tau dt\int d\bsr^o
    \left(\partial_t\calT_Y(\bsr^o,t)\right)^2\po(\bsr^o)\nonumber\\
    &\geq\frac{1}{\mu T\tau}\int d\bsr^o
    \left(\calT_Y^f(\bsr^o)-y^o\right)^2\po(\bsr^o).
\end{align}
In the last inequality, using the method of Lagrange multipliers, the form of the optimal map is specified as the interpolation $\calT(\bsr,t)=(1-t/\tau)\bsr+(t/\tau)\calT^f(\bsr)$.
If we substitute $\calT^f(\bsr)=[\calT_{X|Y}^f(x;y),\calT_Y^{*}(y)]^{\rm T}$ into \eref{apdx: proof of min sigma_Y}, the rightmost of Eq.~(\blue{7}) is obtained.
Here, $\calT_Y^{*}(y)={\Gamma_Y^f}^{-1}(\Gamma_Y^o(y))$ is the one-dimensional optimal map for $Y$.
The $X$ component of optimal transport map, $\calT_{X|Y}^f(x;y)$, that realizes the transport $\po\rightarrow\pf$ is provided by e.g., the Knothe–Rosenblatt map, $\calT_{X|Y}^f(x;y)=\calT_{X|Y}^*(x;y)={\Gamma_{X|Y}^f}^{-1}\left(\Gamma_{X|Y}^o(x;y);\calT_Y^{*}(y)\right)$.
We conclude that Eq.~(\blue{7}) is proved because the marginal Wasserstein distance $\calW(\pyo,\pyf)$ also gives the lower bound of $\sigma_Y$ \cite{Nakazato-Ito2021}.
The Lagrange map for $X$, $\calT_{X|Y}(x,t;y)$, is freely determined as long as continuously connecting $x$ and $\calT_{X|Y}^f(x;y)$ in time.

The above proof does not exclude the existence of a transport map $\calT^f$ that realizes $\po\rightarrow\pf$, where $\calT_Y^f$ depends on $x$ and the minimum EP (Eq.~(\blue{7})) is attained.
We however prove that such a map does not exist in general.
We consider the following Lagrangian:
\begin{align}
    \calL[\calT^f]:=&
    \frac{1}{2}
    \int d\bsr^o\left(\calT_Y^f(\bsr^o)-y^o\right)^2\po(\bsr^o)
    +\int d\bsr^f \lambda(\bsr^f)\left[
    \int d\bsr^o\delta(\bsr^f-\calT^f(\bsr^o))\po(\bsr^o)-\pf(\bsr^f)
    \right],
\end{align}
where $\lambda(\bsr^f)$ represents the Lagrange multiplier.
The variation of the Lagrangian is calculated as
\begin{align}
    \delta\calL=
    \int d\bsr^f\left[\partial_{x^f}\lambda(\bsr^f)\right]\delta\calT_X^f(\calT^o(\bsr^f))\pf(\bsr^f)
    +\int d\bsr^f\left[\partial_{y^f}\lambda(\bsr^f)+y^f-\calT_Y^o(\bsr^f)\right]\delta\calT_Y^f(\calT^o(\bsr^f))\pf(\bsr^f),
\end{align}
where $\calT^o:={\calT^f}^{-1}$ denotes the inverse function of the transport map $\calT^f$. 
From the variation $\delta\calT_X^f(\calT^o(\bsr^f))$, the Lagrange multiplier must be independent of $x$.
Similarly, from the variation $\delta\calT_Y^f(\calT^o(\bsr^f))$, the $Y$ component of the optimal map must also be independent of $x$.

\subsection{Proof of Eq.~(\blue{8})}
We first minimize the EP of $Y$ according to \sref{subsec: proof of min sigma_X}.
The components of the Lagrange map $\calT(\bsr,t)$ satisfy $\calT_Y(\bsr,t)=(1-t/\tau)y+(t/\tau)\calT_Y^*(y)$ and
$\calT_X(\bsr,0)=x,\calT_X(\bsr,\tau)=\calT_{X|Y}(x;y)$.
We next consider minimizing the EP of $X$ under this condition.
The EP of $X$ is 
\begin{align}
    \label{apdx: proof of sigma_X|Y general}
    \sigma_X
    &=\frac{1}{\mu T}\int_0^\tau dt\int d\bsr
    \left(\partial_t\calT_X(\bsr,t)\right)^2\po(\bsr)\nonumber\\
    &\geq\frac{1}{\mu T\tau}\int d\bsr
    \left(\calT_{X|Y}(x;y)-x\right)^2\po(\bsr)\nonumber\\
    &=\frac{1}{\mu T\tau}\int dy \pyo(y)\int dx
    \left(\calT_{X|Y}(x;y)-x\right)^2 p_{X|Y}^o(x|y),
\end{align}
where we used the same reasoning as in \eref{apdx: proof of min sigma_Y} to obtain the inequality in \eref{apdx: proof of sigma_X|Y general}.

\begin{figure}
    \centering
    \includegraphics[width=0.8\linewidth]{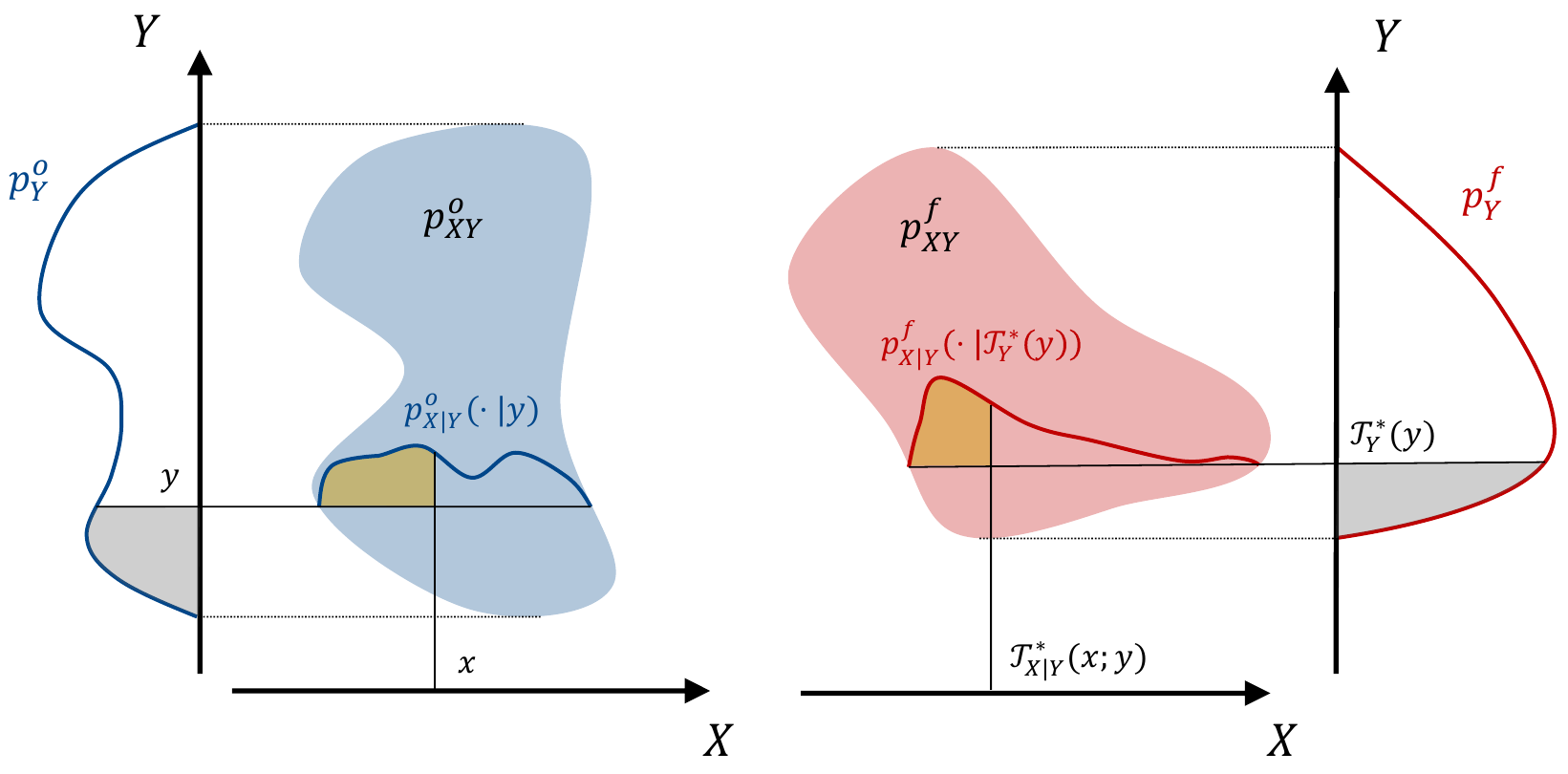}
    \caption{\label{fig: Knothe}
    The optimal map to realize Eq.~(\blue{8}).
    The areas of the shaded regions are identical, respectively for gray and yellow.
    This map is known as Knothe–Rosenblatt map \cite{villani2009}.
    }
\end{figure}

We explain the constraint on the map $\calT_{X|Y}$.
The Jacobian equations (completeness of transport) of $\calT_Y^*(y)$ and $\calT(\bsr)=[\calT_{X|Y}(x;y),\calT_Y^*(y)]^{\rm T}$ are given by
\begin{align}
    \label{apdx: Jacobian Y}
    \pyo(y)&=\frac{\partial\calT_Y^*(y)}{\partial y}\pyf\left(\calT_Y^*(y)\right),
    \\
    \label{apdx: Jacobian XY}
    \po(\bsr)&=\left|\det{\frac{\partial\calT(\bsr)}{\partial \bsr}}\right|\pf\left(\calT(\bsr)\right)
    =\frac{\partial\calT_Y^*(y)}{\partial y}\frac{\partial\calT_{X|Y}(x;y)}{\partial x}
    \pf\left(\calT(\bsr)\right).
\end{align}
We then obtain the Jacobian equation of $\calT_{X|Y}(x;y)$ by dividing \eref{apdx: Jacobian XY} with \eref{apdx: Jacobian Y} as
\begin{align}
    \label{apdx: Jacobian X|Y}
    p_{X|Y}^o(x|y)&=\frac{\partial\calT_{X|Y}(x;y)}{\partial x}p_{X|Y}^f\left(\calT_{X|Y}(x;y)|\calT_Y^*(y)\right).
\end{align}
This indicates that the map $\calT_{X|Y}(\cdot;y)$ realizes the one-dimensional transport from $p_{X|Y}^o(\cdot|y)$ to $p_{X|Y}^f\left(\cdot|\calT_Y^*(y)\right)$.
Combining \eref{apdx: proof of sigma_X|Y general} and \eref{apdx: Jacobian X|Y}, the EP of $X$ is minimized if $p_{X|Y}^o(\cdot|y)$ is transported optimally to $p_{X|Y}^f\left(\cdot|\calT_Y^*(y)\right)$ for all $y$.
For each $y$, the optimal transport cost equals the square of the Wasserstein distance between the conditional distributions, providing Eq.~(\blue{8}).
Since this is the transport of one-dimensional distributions, the optimal transport map is given by Eq.~(\blue{9}).
This map is illustrated in \fref{fig: Knothe}. 

\subsection{Derivation of Eq.~(\blue{12})}
Using Eqs.~(\blue{8})(\blue{11})\eqref{apdx: optimal map Y Gaussian}\eqref{apdx: conditional distribution Gaussian}, $\sigma_X^*|_{\sigma_Y=\sigma_Y^*}$ is calculated as
\begin{align}
    \sigma_X^*|_{\sigma_Y=\sigma_Y^*}
    =&\frac{1}{\mu T\tau}\int dy \pyo(y)
    \left[\left(\left(m_X^o+\frac{\sxy}{\syy}(y-m_Y^o)\right)
    -\left(m_X^f+\frac{\txy}{\tyy}\left(\sqrt{\frac{\tyy}{\syy}}(y-m_Y^o)+m_Y^f-m_Y^f\right)\right)\right)^2
    \right]
    \nonumber\\
    &+\frac{1}{\mu T\tau}\left(\sqrt{\frac{\det{\Sigma^o}}{\syy}}-\sqrt{\frac{\det{\Sigma^f}}{\tyy}}\right)^2\nonumber\\
    &=\frac{1}{\mu T\tau}\left[(m_X^o-m_X^f)^2+\left(\frac{\sxy}{\syy}-\frac{\txy}{\sqrt{\syy\tyy}}\right)^2\syy+
    \left(\sqrt{\frac{\det{\Sigma^o}}{\syy}}-\sqrt{\frac{\det{\Sigma^f}}{\tyy}}\right)^2
    \right]\nonumber\\
    &=\frac{1}{\mu T\tau}\left[\calW(\pxo,\pxf)^2+2\left(\sqrt{\sxx\txx}-\frac{\sxy\txy}{\sqrt{\syy\tyy}}-\sqrt{\frac{\det{\Sigma^o}\det{\Sigma^f}}{\syy\tyy}}\right)
    \right]\nonumber\\
    \label{apdx: proof of sigma_X|Y Gaussian}
    &=\sigma_X^*+\frac{2\sqrt{\sxx\txx}}{\mu T \tau}
    \left(1-\rho_o\rho_f-\sqrt{(1-{\rho_o}^{2})(1-{\rho_f}^{2})}\right).
\end{align}
Thus, Eq.~(\blue{12}) is derived.

%In \eref{apdx: proof of sigma_X|Y Gaussian}, the first term is considered as the energetic dissipation in $X$ required for the local evolution $\po\rightarrow q_{XY}^f$, where $q_{XY}^f$ has the same correlation $\rho_o$ as $\po$ and the same marginal distributions as $\pf$ (see Eq.~(\blue{11})).
%Scaled by $\sqrt{\txx/\sxx}$, the second term corresponds to the energetic dissipation required in $X$ for varying the correlation $\rho_o$ to $\rho_f$ in the time evolution $q_{XY}^f\rightarrow\pf$.
%This EP of $X$ is minimized under the condition of optimizing first the EP of $Y$.
%This understanding is also validated from rewritting \esref{apdx: optimal map XY Gaussian}\eqref{apdx: G1} as
%The matrices $G_{1}^l$ and $ G_{1}^c$ construct the transport map from $\po$ to $q_{XY}^f$ and the transport map from $q_{XY}^f$ to $\pf$, respectively.

%%%%%%%%%%%%%%%%%%%%%%%%%%%%%%%%%%

\section{Pareto front}
In this section, we consider the Pareto front of the partial EPs for overdamped Langevin systems, which is represented by the blue curve of Fig.~\blue{1}.
We note that the Pareto front has also been considered for discrete Markov jump systems in \ccite{kamijima2024discrete}, which however has a very different structure from the Langevin case.

\subsection{Definitions and General Results}

The tradeoff relation between incompatible $\sigma_X$ and $\sigma_Y$ can be described by the set of their optimal pair, called the Pareto front \cite{ngatchou2005pareto,coello2007evolutionary}.
On the Pareto front (denoted as $\calP$), $\sigma_X$ cannot be decreased without increasing $\sigma_Y$, and vice versa (see Fig.~\blue{1}).
Let us define the feasible region $\calK$ as the set of all pairs implemented by some protocols $\{V,F\}_{0\leq t\leq \tau}$ that realize $\po\rightarrow\pf$.
If there exists a feasible pair $(\sigma_X',\sigma_Y')\ (\in\calK)$ such that
\begin{align}
    \sigma_X'<\sigma_X,\ \sigma_Y'\leq\sigma_Y
    \textrm{\ or \ }
    \sigma_X'\leq\sigma_X,\ \sigma_Y'<\sigma_Y,
\end{align}
then the pair $(\sigma_X,\sigma_Y)$ is not included in $\calP$.
This relation between such pairs is denoted as $(\sigma_X',\sigma_Y')\prec(\sigma_X,\sigma_Y)$.
Then, the Pareto front $\calP$ can be more explicitly defined as
\begin{align}
    \calP:=\{(\sigma_X,\sigma_Y)\in\calK\mid
    \forall(\sigma_X',\sigma_Y')\in\calK,
    (\sigma_X',\sigma_Y')\not\prec(\sigma_X,\sigma_Y)\}.
\end{align}
We suppose that $\calK$ is a convex set in $\mathbb{R}^2$, which, however, has not yet been justified in terms of rigorous mathematics.
Then, each point in $\calP$ corresponds to the minimizer of the weighted EP $(1-w)\sigma_X+w\sigma_Y$ for some $w \in(0,1)$ \cite{marler2010weightedsum}.

The minimization of $(1-w)\sigma_X+w\sigma_Y$ can be reduced to an optimal transport problem by squeezing the dynamics (\blue{1}) as follows.
We consider the squeezed coordinate $\tbsr$ as
\begin{align}
    \label{squeezing}
    \tilde{\bsr}=D_w\bsr,
    D_w:=
    \begin{bmatrix}
    \sqrt{1-w} & 0 \\
    0 & \sqrt{w} \\
    \end{bmatrix}.
\end{align}
The probability and mean local velocity are modified by this squeezing as $\tilde{p}_{XY}(\tbsr,t)=(\det{D_w})^{-1}\pt(D_w^{-1}\tbsr,t)$ and $\tbsv(\tbsr,t)=D_w\bsv(D_w^{-1}\tbsr,t)$, respectively.
These satisfy the Fokker-Planck equation $\partial_t \tilde{p}_{XY}=-\tilde{\nabla}^{\rm T}(\tbsv \tpt)$, where $\tilde{\nabla}$ denotes the gradient in the squeezed coordinate.
Since the weighted EP $(1-w){\sigma}_{X}+w{\sigma}_{Y}$ now has the same form as Eq.~(\blue{3}) in terms of $\tpt$ and $\tbsv$, Eq.~(\blue{5}) can be applied to it, yielding
\begin{align}
    \label{min sigma_w general}
    \min_{\substack{\{V,F\}_{0\leq t\leq \tau}\\{\rm s.t.\ }\po\rightarrow\pf}}
    [(1-w){\sigma}_{X}+w{\sigma}_{Y}]=\frac{\calW(\tpo,\tpf)^2}{\mu T\tau}.
\end{align}
The minimum value of the weighted EP is expressed by the Wasserstein distance between the squeezed distributions.
In particular, \eref{min sigma_w general} reduces to Eq.~(\blue{5}) for $w=1/2$ (see Fig.~\blue{1}).

\begin{figure}
    \centering
    \includegraphics[width=0.95\linewidth]{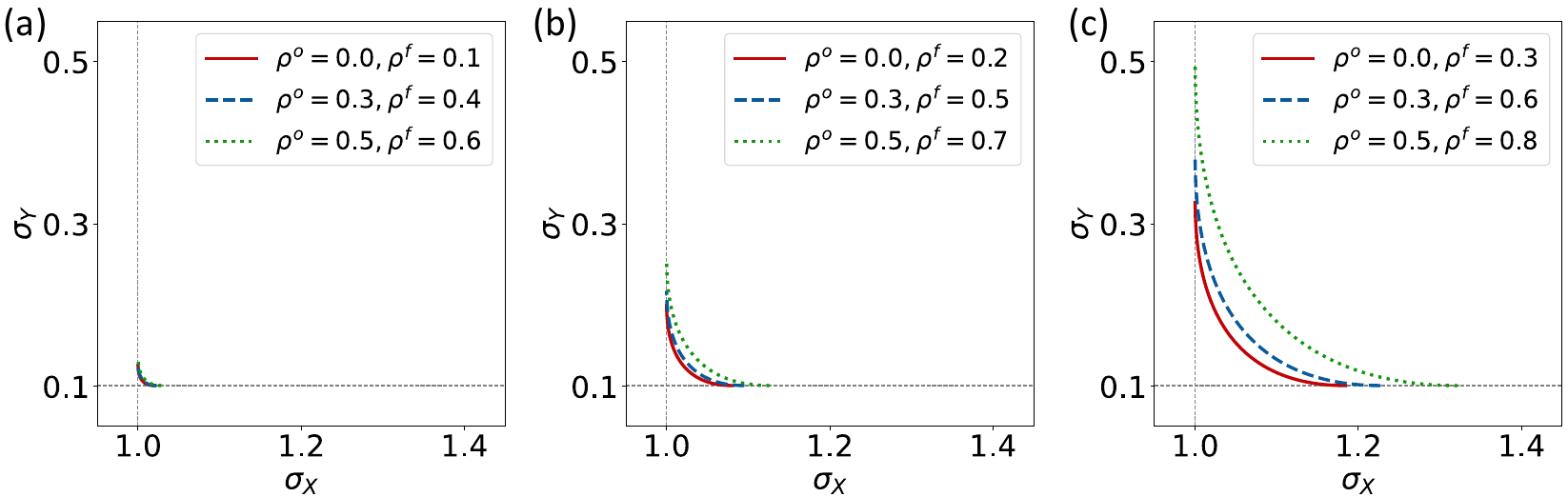}
    \caption{\label{fig: tradeoff_contraction}
    The Pareto fronts of $\sigma_X$ and $\sigma_Y$ for the Gaussian states.
    These fronts are plotted for the different pairs of the correlation coefficients using \esref{apdx: sigma_X w Gaussian}\eqref{apdx: sigma_Y w Gaussian}: (a) $\rho_f-\rho_o=0.1$, (b) $\rho_f-\rho_o=0.2$, and (c) $\rho_f-\rho_o=0.3$.
    The gray dotted lines display $\sigma_X^*$ and $\sigma_Y^*$ (Eq.~(\blue{11})).
    The other parameters are $\sxx=1,\txx=4,\syy=3,\tyy=2,\tau=1$.
    }
\end{figure}

The Pareto front of the partial EPs is obtained by plotting the minimizer of \eref{min sigma_w general} denoted as $(\sigma_X^w,\sigma_Y^w)$ and by sweeping $0<w<1$.
The minimizer and the optimal transport map $\tcalT^{*,w}$ in the squeezed dynamics are uniquely determined \cite{gangbo1996geometry}.
The corresponding transport map in the original dynamics is also unique and expressed as $\calT^{*,w}(\bsr)=D_w^{-1}\tcalT^{*,w}(D_w\bsr)$.
For $w<w'$, $\calT^{*,w}\neq\calT^{*,w'}$  leads to $\sigma_X^w<\sigma_X^{w'},\sigma_Y^w>\sigma_Y^{w'}$, and vice versa.
We note that the Pareto front collapses to a single point when $\calT_X^*=\calT_{X|Y}^*$ holds (see also \fref{fig: tradeoff_contraction}), for example, $X$ and $Y$ are statistically independent in both the initial and final distributions.

From another viewpoint, the minimization \eref{min sigma_w general} is considered as solving an optimal transport problem whose cost is quadratic and weighted, $(1-w)|\calT_X(\bsr)-x|^2+w|\calT_Y(\bsr)-y|^2$.
It is known that the optimal transport map of this problem converges to the Knothe-Rosenblatt map in the limit $w\rightarrow1$ \cite{carlier2010knothe,santambrogio2015OT}, which is expressed as $\lim_{w\rightarrow1}\calT^{*,w}=\calT^{*,1}$.
This also holds for the limit $w\rightarrow0$.
Since the optimal protocol for achieving Eq.~(\blue{7}) is constructed from the map $\calT^{*,1}$, the endpoints of the Pareto front $\calP$ agree with Eqs.~(\blue{7})(\blue{8}):
\begin{align}
    \label{apdx: sigma_X, sigma_Y w1}
    &\lim_{w\rightarrow0}\sigma_X^w=\sigma_X^*,\lim_{w\rightarrow1}\sigma_Y^w=\sigma_Y^*,\\
    \label{apdx: sigma_X, sigma_Y w2}
    &\lim_{w\rightarrow1}\sigma_X^w=\sigma_X^*|_{\sigma_Y=\sigma_Y^*},\lim_{w\rightarrow0}\sigma_Y^w=\sigma_Y^*|_{\sigma_X=\sigma_X^*}.
\end{align}
%（と考えれば(8)の証明はいらないけど、あったほうが断然わかりやすい）

%The variable $\bsv$ of the weighted EP is replaced by the Lagrange map $\calT(\bsr,t)$, which satisfies $\partial_t\calT(\bsr,t)=\bsv(\calT(\bsr,t),t)$ and $\calT(\bsr,0)=\bsr$.
%The conservation of probability is expressed as $\po(\bsr)=|\det(\partial \calT(\bsr,t)/\partial \bsr)|\pt(\calT(\bsr,t),t)$.

%where $\calT(\bsr):=\calT(\bsr,\tau)$ is the transport map and $\parallel\bsa\parallel_w:=((1-w)a_X^2+wa_Y^2)^{\frac{1}{2}}$ is a weighted norm.
%In the third line, we used the fact that the optimal Lagrange map has the form of interpolation, i.e.,  \cite{}.

%We note that the introduced squeezing matrix $D_w$ is quite different from the mobility matrix in \cite{Aurell2012refined} which is canceled out in the protocol.

\subsection{Gaussian Case}
We exemplify the Pareto front $\calP$ for Gaussian distributions.
In this case, the squeezed distributions are also Gaussian and given by
$\pt^{o/f}=\calN(\tilde{\bsm}^{o/f},\tilde{\Sigma}^{o/f})$
with $\tilde{\bsm}^{o/f}:=D_w\bsm^{o/f},\tilde{\Sigma}^{o/f}:=D_w\Sigma^{o/f} D_w$.
The Wasserstein distance between these Gaussian distributions is calculated analytically as \cite{Givens1984Wasserstein-Gaussian}
\begin{align}
    \label{Wasserstein Gaussian squeezed}
    \calW(\tpo,\tpf)^2
    =\parallel \tilde{\bsm}^o-\tilde{\bsm}^f\parallel^2
    +\trace{\tilde{\Sigma}^o}+\trace{\tilde{\Sigma}^f}-2\trace{\left((\tilde{\Sigma}^o)^{\frac{1}{2}}\tilde{\Sigma}^f(\tilde{\Sigma}^o)^{\frac{1}{2}}\right)^{\frac{1}{2}}},
\end{align}
where $A^{1/2}$ denotes the square root of the positive definite symmetric matrix $A$.
The optimal transport map $\tcalT^{*,w}$ is linear
$\tcalT^{*,w}(\tbsr)=G(\tilde{\Sigma}^{o-1},\tilde{\Sigma}^f)(\tbsr-\tbsm^o)+\tbsm^f$, where $G(A,B):=A^{\frac{1}{2}}(A^{-\frac{1}{2}}BA^{-\frac{1}{2}})^{\frac{1}{2}}A^{\frac{1}{2}}$ is the geometric mean of positive definite symmetric matrices $A,B$ \cite{bhatia2019bures}.
Accordingly, the corresponding transport map in the original dynamics is also linear:
\begin{align}
    \calT^{*,w}(\bsr)=D_w^{-1}G(\tilde{\Sigma}^{o-1},\tilde{\Sigma}^f)D_w(\bsr-\bsm^o)+\bsm^f=:G_w(\bsr-\bsm^o)+\bsm^f.
\end{align}
From \eref{apdx: proof of min sigma_Y}, the partial EPs of the corresponding protocol are calculated as
\begin{align}
    \label{apdx: sigma_X w Gaussian}
    &\sigma_X^w=\frac{1}{\mu T\tau}\left[(m^o_X-m^f_X)^2+\sxx+\txx-2\trace{(D_0G_w\Sigma^o)}\right],\\
    \label{apdx: sigma_Y w Gaussian}
    &\sigma_Y^w=\frac{1}{\mu T\tau}\left[(m^o_Y-m^f_Y)^2+\syy+\tyy-2\trace{(D_1G_w\Sigma^o)}\right].
\end{align}
%Therefore, the Pareto front $\calP$ is obtained by plotting $(\sigma_X^w,\sigma_Y^w)$ for all $0<w<1$.
The blue curve in Fig.~\blue{3}(d) is obtained by plotting these EPs for $0<w<1$.
The Pareto front contracts to a single point if and only if $\rho_o=\rho_f$ (see Eqs.~(\blue{12})\eqref{apdx: optimal map XY Gaussian}).
The Pareto front expands further as $|\rho_o-\rho_f|$ increases (see \fref{fig: tradeoff_contraction}).

We examine the limits of $w\rightarrow0$ and $w\rightarrow1$ to validate \esref{apdx: sigma_X, sigma_Y w1}\eqref{apdx: sigma_X, sigma_Y w2}.
Using the relation $\trace{A^{\frac{1}{2}}}=\sqrt{\trace{A}+2\det{A}^{\frac{1}{2}}}$ for the positive definite symmetric matrix $A$ of size $2$, it can be shown that the limits of these EPs satisfy \eref{apdx: sigma_X, sigma_Y w1},
where $\sigma_X^*,\sigma_Y^*$ are given by Eq.~(\blue{11}).
Furthermore, one can confirm that the limit of $G_w$ agrees with \eqref{apdx: G1}, that is, $\lim_{w\rightarrow1}=G_1$ from the asymptotic analysis.
This also holds for the limit $w\rightarrow0$.
Thus, \eref{apdx: sigma_X, sigma_Y w2} is verified through the calculation of \esref{apdx: sigma_X w Gaussian}\eqref{apdx: sigma_Y w Gaussian}, where $\sigma_X^*|_{\sigma_Y=\sigma_Y^*},\sigma_Y^*|_{\sigma_X=\sigma_X^*}$ are given by Eq.~(\blue{11}).

%The result \eqref{min sigma_Y general} can be understood as follows.
%The probability is squeezed to the $X$ axis for small $w\ll1$ and the protocol in terms of $X$ thus approaches the one-dimensional optimal transport of the marginal distribution (see \fref{fig: squeezing}).
%We note that the complete squeezing (projection) $w=0$ leads to the loss of information about the subsystem $Y$ and the unsqueezing cannot extract the optimal protocol in the original dynamics.

%ちょっと一般化できるf(x)[Amari]

%\bibliography{ref}
%

%%%%%%%%%%%%%%%%%%%%%%%%%%%%%%%%%%

%% file: commands.tex
\newcommand{\todayd}{\the\year/\the\month/\the\day}

\newcommand{\bel}{\begin{easylist}}
\newcommand{\eel}{\end{easylist}}
\newcommand{\eref}[1]{Eq.~\eqref{#1}}
\newcommand{\esref}[1]{Eqs.~\eqref{#1}}

\newcommand{\fref}[1]{Fig.~\ref{#1}}
\newcommand{\fsref}[1]{Figs.~\ref{#1}}

\newcommand{\ccite}[1]{Ref.~\cite{#1}}
\newcommand{\scite}[1]{Refs.~\cite{#1}}

\newcommand{\sumtwo}[2]%
{\mathop{\sum_{#1}}_{#2}}
\newcommand{\sumthree}[3]%
{\mathop{\mathop{\sum_{#1}}_{#2}}_{#3}}
\newcommand{\sumfour}[4]%
{\mathop{\mathop{\mathop{\sum_{#1}}_{#2}}_{#3}}_{#4}} 
\newcommand{\prodtwo}[2]%
{\mathop{\prod_{#1}}_{#2}}
\newcommand{\mintwo}[2]%
{\mathop{\min_{#1}}_{#2}}
\newcommand{\maxtwo}[2]%
{\mathop{\max_{#1}}_{#2}}
\newcommand{\maxthree}[3]%
{\mathop{\mathop{\max_{#1}}_{#2}}_{#3}}
\newcommand{\limtwo}[2]%
{\mathop{\lim_{#1}}_{#2}}
\newcommand{\suptwo}[2]%
{\mathop{\sup_{#1}}_{#2}}
\newcommand{\supthree}[3]%
{\mathop{\mathop{\sup_{#1}}_{#2}}_{#3}}
\newcommand{\supfour}[4]%
{\mathop{\mathop{\mathop{\sup_{#1}}_{#2}}_{#3}}_{#4}} 
\newcommand{\inftwo}[2]%
{\mathop{\inf_{#1}}_{#2}}
\newcommand{\infthree}[3]%
{\mathop{\mathop{\inf_{#1}}_{#2}}_{#3}}
\newcommand{\inffour}[4]%
{\mathop{\mathop{\mathop{\inf_{#1}}_{#2}}_{#3}}_{#4}} 
\newcommand\calF{{\mathcal F}}

\newcommand\calN{{\mathcal N}}

\newcommand\calT{{\mathcal T}}

\newcommand\calW{{\mathcal W}}

\newcommand{\bsm}{\boldsymbol{m}}

\newcommand{\bsr}{\boldsymbol{r}}

\newcommand{\bsv}{\boldsymbol{v}}

\newcommand{\bsF}{\boldsymbol{F}}

\newcommand{\pt}{p_{XY}}
\newcommand{\po}{p^o_{XY}}
\newcommand{\pf}{p^f_{XY}}

\newcommand{\pyo}{p^o_{Y}}
\newcommand{\pyf}{p^f_{Y}}

\newcommand{\sxx}{\Sigma^o_{XX}}

\newcommand{\syy}{\Sigma^o_{YY}}
\newcommand{\txx}{\Sigma^f_{XX}}

\newcommand{\tyy}{\Sigma^f_{YY}}